\documentstyle[12pt,aaspp,epsf]{article}
\def\v4{$V_4$}
\def\i4{$I_4$}
\def\asec{\rlap{$^{\prime\prime}$}.\hbox to 2pt{}}
\def\amin{\rlap{$^\prime$}.\hbox to 1pt{}}
\def\mag{\hskip 0.5pt\rlap{$^{\rm m}$}{\hskip 1.5pt.}\hbox to 2.0pt{}}

\def\square{\vbox to 7pt{\boxit{\hbox to 0pt{}}}\hskip 1pt}

\def\footnoterule{\kern-3pt\hrule\kern 2.6pt}

\newbox\grsign \setbox\grsign=\hbox{$>$} \newdimen\grdimen \grdimen=\ht\grsign
\newbox\simlessbox \newbox\simgreatbox
\setbox\simgreatbox=\hbox{\raise.5ex\hbox{$>$}\llap
     {\lower.5ex\hbox{$\sim$}}}\ht1=\grdimen\dp1=0pt
\setbox\simlessbox=\hbox{\raise.5ex\hbox{$<$}\llap
     {\lower.5ex\hbox{$\sim$}}}\ht2=\grdimen\dp2=0pt
\def\simgreat{\mathrel{\copy\simgreatbox}}
\def\simless{\mathrel{\copy\simlessbox}}

\def\vol#1  {{{#1}{\rm,}\ }}

\def\aj{{AJ}, }  
\def\apj{{ApJ}, } 
\def\apjs{{ApJS}, } 
\def\apjl{{ApJ}, } 
\def\mnras{{MNRAS}, } 
\def\aasup{{A\&AS}, } 
\def\hi{\noindent \hangindent=2.5em}

\def\clock{\count0=\time \divide\count0 by 60
     \count1=\count0 \multiply\count1 by -60 \advance\count1 by \time
     \number\count0:\ifnum\count1<10{0\number\count1}\else\number\count1\fi}

\begin{document}

\title{ The Palomar Distant Cluster Survey :  \\
        III. The Colors of the Cluster Galaxies}

\author{Lori M. Lubin\altaffilmark{1}}
\affil{Princeton University Observatory, Peyton Hall, Princeton NJ 08544}
\affil{lml@ociw.edu}

\vskip 3 cm
\centerline{Accepted for publication in the {\it Astronomical Journal}}


\altaffiltext{1}{Present Address : Observatories of the Carnegie Institution
of Washington, 813 Santa Barbara Street, Pasadena, CA 91101}

\vfill
\eject

%
%

\begin{abstract}
We present a color analysis of the galaxy populations of candidate
clusters of galaxies from the Palomar Distant Cluster Survey (Postman
et al.\ 1996).  The survey was conducted in two broad band filters
that closely match $V$ and $I$ and contains a total of 79 candidate
clusters of galaxies, covering an estimated redshift range $0.2
\simless z \simless 1.2$.  We examine the color evolution in the 57
richest clusters from this survey, the largest statistical sample of
distant clusters to date. The intermediate redshift ($0.2 \simless z
\simless 0.4$) clusters show a distinct locus of galaxy colors in the
color--magnitude diagram. This ridge line corresponds well with the
expected no--evolution color of present--day elliptical galaxies at
these redshifts. In clusters at redshifts of $z \simgreat 0.5$, this
red envelope has shifted bluewards compared to the ``no--evolution''
prediction. By $z \sim 0.8$ there are only a few galaxies which are as
red in their rest-frame as present--day ellipticals, consistent with
recent claims on the basis of optical--infrared colors. The detected
evolution is consistent with passive aging of stellar populations
formed at redshifts of $z \simgreat 2$.

Though the uncertainties are large, the Butcher--Oemler effect is
observed in the Palomar clusters. The fraction of blue galaxies
increases with the estimated redshift of the cluster at a 96.2\%
confidence level. The measured blue fractions of the intermediate
redshift clusters ($f_{b} \sim 0.2 - 0.3$) are consistent with those
found previously by Butcher \& Oemler (1984). The trend in the Palomar
clusters suggests that $f_{b}$ can be greater than 0.4 in clusters of
galaxies at redshifts of $z \simgreat 0.6$.

\end{abstract}

{\it Subject headings}: galaxies: clusters of galaxies; cosmology:
observations; evolution

\section{Introduction}

The study of the galaxy populations of rich clusters provides
important constraints on the formation mechanisms of both clusters and
galaxies. Present--day clusters show a distinct correlation between
the structure of the cluster and the galaxy population. Irregular,
open clusters, such as Virgo, are spiral--rich. These systems show no
obvious condensations, though the galaxy surface density is at least
five times as great as the surrounding field ($n_{\rm gal} >
30~h^{3}~{\rm galaxies~{Mpc}^{-3}}$). These clusters may be highly
assymetric and have significant degrees of substructure.  Dense,
centrally concentrated clusters, such as Coma, contain predominantly
early--type galaxies in their cores (Abell 1958; Oemler 1974; Dressler
1980; Postman \& Geller 1984).  These clusters have a single,
outstanding concentration among the bright member galaxies and
typically display a high--degree of spherical symmetry. Central
densities can reach as high as $10^{4}~h^{3}~{\rm
galaxies~{Mpc}^{-3}}$. The galaxy content of clusters is part of the
general morphology--density relation of galaxies; as the local density
increases, the fraction of elliptical (E) and S0 galaxies increases,
while the fraction of spiral galaxies decreases (Dressler 1980;
Postman \& Geller 1984).

One of the most intriguing results of the study of intermediate and
high--redshift clusters of galaxies has been strong evolution in the
galaxy population.  Butcher \& Oemler (1984; hereafter BO) were the
first to make a comprehensive study of intermediate redshift clusters
with regard to their galaxy populations. They examined the fraction of
blue galaxies ($f_{b}$) in 33 clusters at $z \simless 0.5$. They found
that $f_{b}$ is an increasing function of redshift in both open and
compact clusters of galaxies, indicating that clusters at these
redshifts are significantly bluer than their low--redshift
counterparts. Recent HST image data (Dressler et al.\ 1994; Couch et
al.\ 1994; Oemler, Dressler \& Butcher 1996) reveal that many of these
blue ($g-r < 1.2$) galaxies are either ``normal'' spirals or have
peculiar morphologies, producing non--elliptical fractions which are 3
to 5 times higher than the average current epoch cluster.  Fading
through cessation of star formation may play a role in this
evolution. Gunn \& Dressler (1988) find that the spectra of cluster
galaxies with $z \simgreat 0.6$ show, on average, smaller 4000$\AA$
decrements and a higher frequency of post--starburst features (the
``E+A" spectral class) than those at $z < 0.6$ (however, see Zabludoff
et al.\ 1996).

Detailed photometric observations of other intermediate redshift ($z
\simless 0.4$) clusters have confirmed the original results of BO.
Even though these clusters show an increased fraction of blue
galaxies, they still contain a population of E/S0s which distinguish
itself by extremely red colors and a tight color--magnitude (CM)
relation (a ``red envelope'').  Both the mean color and the CM
relation is consistent with that of present--day ellipticals (Sandage
1972; Visvanathan \& Sandage 1977; Butcher \& Oemler 1978; Couch \&
Newell 1984; BO; Ellis et al.\ 1985; Sandage, Bingelli \& Tammann
1985; Millington \& Peach 1990; Arag\'on-Salamanca, Ellis \& Sharples
1991; Luppino et al.\ 1991; Dressler \& Gunn 1992; Le Borgne, Pell\'o
\& Sanahuja 1992; Dressler et al.\ 1994; Molinari et al.\ 1994; Smail,
Ellis \& Fitchett 1994; Stanford, Eisenhardt \& Dickinson 1995).

Arag\'on-Salamanca et al.\ (1993; hereafter A93) have studied a small
sample of 10 rich clusters at $0.5 < z < 0.9$ in the optical--infrared
colors.  They observe an increase in the number of blue members which
they interpret as the high--redshift extension to the Butcher--Oemler
effect. This trend is also well studied by Rakos \& Schombert (1995)
in an intermediate and high--redshift sample of 17 rich clusters. They
find the fraction of blue galaxies increases from 20\% at $z = 0.4$ to
80\% at $z = 0.8$ using Str\"omgren photometry in the cluster
rest-frame.  In addition, the red envelope, presumably that of the
early--type population, moves {\it bluewards} with redshift (A93;
Rakos \& Schombert 1995; Oke, Gunn \& Hoessel 1996). At $z \sim 0.9$,
there are few cluster members with colors as red as present--day
ellipticals (see also Smail et al.\ 1994). The color distribution of
this high-redshift elliptical population is relatively narrow, and the
trend is uniform from cluster to cluster; this suggests a homogeneous
population which formed within a narrow time span (e.g.\ Bower, Lucey
\& Ellis 1992a,b). Dickinson (1995) finds similar results in a cluster
of galaxies which is associated with the $z = 1.206$ radio galaxy 3C
324. The galaxies exhibit a narrow, red locus in the CM magnitude
diagram. This branch is $\sim 0.6$ mag bluer than the expected
``no--evolution'' value, though the intrinsic rms color scatter is
only 0.2 mag. The observed color trend for the red envelope of
galaxies is consistent with passive evolution of an old stellar
population formed by a single burst of star formation at redshifts of
$z \simgreat 2$.  The reasonably small color scatter would imply
closely synchronized intra--cluster star formation (Bower et al.\
1992a,b; A93; Dickinson 1995).

It is critical to test the claims of significant color evolution in
high--redshift clusters of galaxies against larger, more statistically
complete samples. In this paper, we examine the galaxy colors of the
largest, statistically complete sample of distant clusters presently
available, the Palomar Distant Cluster Survey (hereafter PDCS; Postman
et al.\ 1996). This sample of 57 distant clusters will be used to
trace the color evolution in the galaxy population and to investigate
the Butcher--Oemler effect in clusters up to a redshift of $\sim
1$. In \S 2 of this paper, we briefly describe the PDCS, the original
photometric reduction, and the cluster sample used in this analysis.
The evolution in the color distributions of the cluster galaxies is
presented in \S 3. The Butcher--Oemler effect (the fraction of blue
galaxies as a function of redshift) is presented in \S 4. We summarize
the results in \S 5.

\section{The Data}

The observations, data reduction, and cluster catalog are the subject
of the first paper in this series (Postman et al. 1996; hereafter Paper
I).  However, we discuss briefly the aspects of the original survey
which are necessary for the following analysis. The cluster sample is
derived from an optical/near IR survey with the 4--shooter CCD camera
on the Palomar 5 meter telescope. The survey covers five fields, each
of which is approximately one degree square. The five regions will be
denoted as the ${00}^{{\rm h}}$, ${02}^{{\rm h}}$, ${09}^{{\rm h}}$,
${13}^{{\rm h}}$, and ${16}^{{\rm h}}$ fields.

\subsection{The Photometry}

The Palomar Distant Cluster Survey was conducted in two broad band
filters, the F555W and F785LP of HST's Wide Field/Planetary Camera. We
denote these bands \v4 (F555W) and \i4 (F785LP) according to the
convention in Paper I.  The response curves of these filters are shown
in Fig.\ 1 of Paper I. The rough transformations to the $V$ band and
the Kron-Cousins $I_{KC}$ band are given by

\begin{eqnarray}
V & = & V_{4} - 0.02 - 0.056~(V_{4} - I_{4}) + 0.012~{(V_{4} - I_{4})}^2 \\
I_{KC} & = & I_{4} - 0.43 + 0.089~(V_{4} - I_{4})
\end{eqnarray}

\noindent Photometric conversions to other standard photometric
systems are given in Paper I. The zero points of the \v4 and \i4
magnitudes are based on the AB magnitude system of Oke \& Gunn (1983).
The data are complete to roughly $V_{4}^{iso} = 23.8$ and $I_{4}^{iso}
= 22.5$ in the isophotal magnitudes. The estimated uncertainties in
the FOCAS photometry are $\pm 0.12$ mag for a galaxy with $V_{4}^{iso}
= 22.0$ and $I_{4}^{iso} = 21.0$. The zero point fluctuations from
field to field are estimated at $\simless 0.07$ mag (see Paper I for
details).

The object detection and classification was performed with a modified
version of FOCAS (Jarvis \& Tyson 1981; Valdes 1982).  The basic FOCAS
object detection, point spread function (PSF), measurement,
deblending, and star--galaxy--noise classification algorithms are
used. The \v4 and \i4 band CCD images are analyzed individually.  The
object isophotal detection threshold in each frame is set to be
$3\sigma_{sky}$ (typically 25.7 mag per ${\rm arcsec^{2}}$ in the \v4
band and 24.8 mag per ${\rm arcsec^{2}}$ in the \i4 band), and a
minimum object size requirement of 15 pixels (1.68 ${\rm arcsec^{2}}$)
is imposed.  The sky background is defined as the mode of values in a
given region surrounding the isophotal limits of the galaxy. Because
the sky level is lower, and the system response is higher in the \v4
band than in the \i4 band, the \v4 images are deeper.  Consequently,
about 40\% of the objects detected in a typical \v4 image are
undetected by FOCAS in the corresponding \i4 image.  Nearly all of
these objects are faint (isophotal magnitude of $V_{4}^{iso} \simgreat
23$). About 95\% of the objects with $\simless 23$ are detected in
both bands.

In the following analysis, we use those objects which have been
classified as galaxies by the FOCAS algorithm in the band where the
object is brightest. The object classification accuracy is discussed
extensively in \S 3.3 of Paper I. The classifier breaks down for
objects fainter than about 1 magnitude above the completeness
limit. Near the limit, the signal-to-noise is low ($S/N \sim 4$), and
the mean object area is $\simless 5 \times {\rm FWHM^{2}}$, making
reliable classification difficult. This effect may mean that some
distant ellipticals (those galaxies that we specifically want to
examine) are misclassified as stars. However, because galaxies
outnumber stars by about 7 to 1 at the magnitude limit of the sample,
this is not a significant effect (see Figs.\ 7 \& 8 of Paper I). In
addition, we have confirmed that the results of this paper remain
unchanged even if we had included the stars in the color analysis and
simply subtracted them through the statistical field subtraction (see
\S 3).

For the color analysis in this paper, we have obtained aperture
photometry on all galaxies detected individually in the \v4 and \i4
bands as described above. A circular aperture of 5.03 arcsec in radius
was used.  This corresponds to a physical size of
$\{13.8~19.1~21.1\}~h^{-1}~{\rm kpc}$ at $z = \{0.3~0.6~0.9\}$. The
aperture magnitude limits are $V_{4}^{ap} \approx 22.8$ and
$I_{4}^{ap} \approx 21.4$. The uncertainty in the ${(V_{4} -
I_{4})_{ap}}$ color is estimated at $\sim 0.2$ mag for a galaxy close
to our aperture magnitude limits. In the analysis which follows, we
assume $q_{o} = 0.5~{\rm and}~H_{o} = 100~h~{\rm
km~s^{-1}~Mpc^{-1}}~{\rm with}~h = 0.75$.

\subsection{The Cluster Sample}

A matched filter algorithm was used to objectively identify the
cluster candidates by using positional and photometric data
simultaneously.  This technique is likely to be more robust than
previous optical selections which simply looked for surface density
enhancements, a method which can be significantly affected by
superposition effects (e.g.\ Abell 1958; Gunn, Hoessel \& Oke 1986;
Couch et al.\ 1991). An advantage of this technique is that redshift
estimates of the cluster candidates are produced as a byproduct of the
matched filter; the main disadvantage is that we must assume a
particular form for the cluster luminosity function (for the flux
filter) and cluster radial profile (for the radial filter).  The
radial filter $P(r)$ and the flux filter $L(m)$ are given by
 
\begin{eqnarray}
P(r) = & {1 \over {\sqrt{1 + (r/r_{c})^{2}}}} - {1 \over {\sqrt{1 + (r_{co}/r_c)^{2}}}} & \mbox{if $r < r_{co}$} \nonumber \\
        & 0 & \mbox{              otherwise}
\end{eqnarray}
 
\begin{equation}
L(m) = {{\phi(m-m^{*})~10^{-0.4(m-m^{*})}}\over{b(m)}}
\end{equation}
 
\noindent $P(r)$ is an azimuthally symmetric cluster surface density
profile which has a characteristic core radius ($r_{c}$) and which
falls off at large radii as $r^{-1}$.  The function is explicitly cut
off at an arbitrary cutoff radius ($r_{co}$). We have chosen $r_{c} =
100~h^{-1}~{\rm kpc}$ and $r_{co} = 10 \times r_{c}$ such that the
radial profile resembles the profiles of nearby clusters and that we
have optimized the cluster detections relative to the spurious
detection rates (Paper I). $b(m)$ is the background galaxy
counts. $\phi(m-m^{*})$ is the differential Schechter luminosity
function with $\alpha = -1.1$ and $M^{*} = -21.0$ and $-21.9$ in the
\v4 and \i4 bands, respectively; here, we assume the shape of the
luminosity function is independent of redshift and adopt a
k--correction appropriate for a {\it non--evolving elliptical} galaxy
(see Fig.\ 1). For the derivation of these filters and a detailed
explanation, see Paper I. We have used extensive Monte--Carlo
simulations in Paper I and Lubin \& Postman (1996; hereafter Paper II)
to quantify the selection function due to the functional form of the
matched filter. We find that the selection bias has a minimal effect
on the properties of the observed clusters. We can detect clusters
with a broad range of profile shapes, luminosity function parameters,
and color evolution (Paper I; Paper II).

The catalog consists of 79 candidate clusters of galaxies detected
with estimated redshifts between $0.2 \le z_{est} \le 1.2$.  The
estimated redshifts are determined at discrete 0.1 intervals and
represent the redshift at which the cluster candidate best matches the
filter.  The uncertainty in the estimated redshift ($z_{est}$) is
$\sigma_{z_{est}} \simless 0.2$. This value is determined from both
the simulations and those clusters in the survey which have measured
redshifts. Presently, only 10\% of the cluster sample have reliably
measured redshifts, although they span almost the entire range in
redshift, $0.3 < z_{obs} < 1.1$.  In all cases, the estimated
redshifts are within $z_{obs} \pm 0.2$ (see Fig.\ 15 and \S 4.2.1 of
Paper I).  The redshift survey of the Palomar clusters is ongoing, and
those interested in its current status should contact the authors.

Candidate clusters are detected individually in each band and then
matched with the other to locate those systems which are detected in
both bands.  87\% of the cluster candidates are matched detections;
that is, they have been significantly detected in both the \v4 and \i4
bands. The amplitude of the matched filter provides an estimate of the
cluster richness. This filter richness ($\Lambda_{cl}$) is a measure
of the effective number of $L^{*}$ galaxies in the cluster (see \S
4.2.2 of Paper I; Lubin 1995). Through Monte--Carlo simulations, we
can statistically determine the relation between $\Lambda_{cl}$ and
the actual cluster richness as determined by the specification of
Abell (1958).  This relation is dependent on profile shape; as the
cluster profile slope steepens, a given $\Lambda_{cl}$ value
corresponds to a {\it lower} richness class. For a cluster which has a
surface density profile of approximately $r^{-1.4}$ (the average
profile of the PDCS clusters; Paper II), $\Lambda_{cl} \simgreat 40$
corresponds to Abell ${\rm R} \ge 1$ (Paper I).  Therefore, in the
analysis which follows, we will examine only those clusters which have
been significantly detected in both bands with $\Lambda_{cl} \ge 40$
in either of the two bands. This corresponds to a sample of 57
clusters.  We list in Table 1 the cluster ID \# and the estimated
redshift ($z_{est}$) and filter richness parameter ($\Lambda_{cl}$) as
determined in the passband where the significance of the detection is
highest (see Paper I).

\section{Color Evolution}

In the following analysis, we use the color--magnitude (CM) diagram
and the color distribution to examine the color evolution in the PDCS
cluster galaxies as a function of redshift. Though it is difficult to
select a morphological class based solely on a single broad--band
color, we have tried to identify the high--redshift counterparts to
nearby ellipticals and S0s by looking for their distinct locus or
``red envelope'' in the CM diagram of clusters. At $z \simless 0.4$,
the E/S0 population in clusters has colors which are negligibly
different from their present--day counterparts (see \S 1). At higher
redshifts, there is a systematic and monotomic evolution as a function
of redshift in the optical-infrared colors of early-type galaxies in
clusters (A93; Rakos \& Schombert 1995; Oke et al.\ 1996). At $z \sim
0.9$ there are few cluster galaxies as red as present--day ellipticals
(A93; Dickinson 1995; Rakos \& Schombert 1995).

We examine these trends in the \v4 and \i4 bands of the PDCS sample of
intermediate and high redshift clusters of galaxies.  On an individual
cluster basis, we are limited by statistics and the uncertainty in the
estimated redshift ($z_{est}$); therefore, we follow the same approach
as in Paper II and create global cluster composites for our color
analysis.  Because the cluster redshifts have an estimated uncertainty
of $\sigma_{z_{est}} \simless 0.2$ (see \S2.2 and \S4.2.1 of Paper I),
we sort the cluster sample by estimated redshift into three broad
redshift ranges : (1) $0.2 \le z_{est} \le 0.4$, (2) $0.5 \le z_{est}
\le 0.7$, and (3) $0.8 \le z_{est} \le 1.2$. $z_{est}$ is determined
in discrete 0.1 redshift intervals (see Paper I). Table 2 lists the
number of PDCS clusters per redshift interval in each of the five
fields.

In examining the color data, we adhere to the convention of displaying
the observed colors for the galaxies in our intermediate and
high--redshift cluster samples; that is, we do not reduce our data to
their zero--redshift equivalent by applying k--corrections. In making
comparisons to present--day populations of cluster galaxies, we have
dimmed appropriately the current epoch data (see \S 3.1).

\subsection{No Evolution Predictions}

In order to compare our observations with those at zero redshift, we
calculate the color evolution of typical nearby galaxies of various
morphological type assuming ``no evolution.'' We have used the
spectral energy distributions (SEDs) from Coleman, Wu \& Weedman
(1980) and calculated the \v4 and \i4 k--corrections and expected
$V_{4} - I_{4}$ colors as a function of redshift and morphological
type (see also Frei \& Gunn 1994; Fukugita, Shimasaku \& Ichikawa
1995).  Fig.\ 1 shows the results of these calculations. The
correction to the galaxy color due to the relative change in projected
linear size of the photometric aperture is small compared to other
uncertainties ($\simless 0.05~{\rm mag}$; Sandage \& Visvanathan 1978;
Silva \& Elston 1994).

Because we are trying to assess the color evolution in the cluster
galaxies over a broad range in redshift, we need to quantify our
selection biases due to the magnitude limit of our survey.  Therefore,
we have created a {\it no--evolution} simulation by examining a
``synthetic'' cluster population over our redshift range.  We have
constructed this population from CL 0939+4713 ($z = 0.41$) which has
galaxies with measured colors and morphological classifications from
HST (Dressler et al. 1994).  Dressler et al.\ (1994) found that this
cluster exhibited a blue fraction consistent with that expected from
the Butcher--Oemler effect. The colors of the cluster galaxies suggest
that most of them are normal examples of their Hubble type, and their
luminosity functions are consistent with that of present--day cluster
galaxies. We have converted the $g$ and $r$ magnitudes of Dressler \&
Gunn (1992) to the \v4 and \i4 bands (see Paper I for photometric
conversions). Fig.\ 2 shows the color--magnitude diagram and color
distribution of this simulated no-evolution galaxy population as a
function of redshift. We have chosen the redshifts $z =
\{0.3~0.6~0.9\}$ because they are roughly the medians of the 3
estimated redshift intervals being studied here.  When calculating the
aperture magnitude ($I_{4}^{ap}$) at these redshifts, we have taken
into account the change in projected linear size of the photometric
aperture with redshift; however, we have not corrected the $V_{4} -
I_{4}$ colors for the effect of a color gradient as the correction is
small (see Sandage \& Visvanathan 1978; Silva \& Elston 1994).

The E/S0s (filled circles) appear on a distinct locus in each of these
diagrams. Fig.\ 2 also shows the total color distribution (solid line
histogram) and the distribution that we would be observed given the
magnitude limits of our survey (shaded histogram). In order to compare
the actual data with these evolutionary predictions, we need a
quantitative description of these color distributions. We would like
an estimator of the mean color (presumably that of the early--type
population) which is not severely affected by the presence of a blue
tail. We follow the approach used in A93 and adopt the biweight
location ($C_{BI}$) and scale ($S_{BI}$) estimators (Beers, Flynn \&
Gebhardt 1990 and references therein).  These estimators are
well-suited to non-Gaussian distributions.  For a Gaussian
distribution, $C_{BI}$ is better than 80\% efficient at accurately
determining the mean with more than 10 points, while $S_{BI}$
asymptotically approaches the standard deviation (see Beers et al.\
1990 for exact details). In order to gauge the $1 \sigma$ errors on
these parameters, we have used a bootstrap analysis. This technique
involves randomly drawing many new samples (of equal number) from the
original sample of galaxy colors. The standard deviations of the
resulting parameter distributions are defined as the $1\sigma$
parameter errors. Table 3 lists these parameters for the total and
observed color distributions of the simulated cluster populations. The
more conventional descriptors, the median and interquartile range,
give results which are similar to those of the biweight location and
scale estimators.

Fig.\ 2 shows that, at $z = 0.3$, the simulated color distributions
are unaffected by the magnitude limit of the survey.  At $z = 0.6$,
the E/S0 population still appears as a clear peak in the color
distribution (\v4 -- \i4 $\sim 2.2$), even though a large portion of
the cluster population extends beyond the magnitude limits of the
survey.  At $z \sim 0.9$, the observed color distribution (bottom
panels of Fig.\ 2) is significantly altered, though it would still be
possible to detect the brightest cluster ellipticals at $I_{4}^{ap}
\simless 20$ with a predicted color of ${(V_{4} - I_{4})}_{ap} \sim
3$. The mean color in this redshift interval is $C_{BI} \sim
2.4$. Note that the differential k-correction (Fig.\ 1) causes a
distinct change in the relative number of E/S0s versus spirals
observed in the different redshift bins. For the chosen synthetic
cluster population, the ratio of E/S0s to spirals is $\sim 3$; this
ratio is preserved in the observed color distribution of a cluster at
$z \sim 0.3$ but becomes $\sim 1$ for a cluster at $z \sim 0.9$.

For the analysis of the PDCS clusters, we study {\it composite} color
distributions (see \S 3.2).  Since we co-add clusters over a broad
range in redshifts, this will artificially increase the dispersion
around the elliptical color--magnitude relation, as well as the width
of the color distribution in general.  For example, from $z = 0.2
\rightarrow 0.4$ the no--evolution elliptical color changes by $\sim$
0.6 mag (Fig.\ 1).  We simulate this effect by generating the
synthetic cluster population of Fig.\ 2 at random redshifts between
$0.1 \simless z \simless 1.2$.  We then create composite color
diagrams for each redshift interval such that we have roughly the same
number of galaxies as observed in the PDCS composites (see Fig.\ 7).
These, therefore, are the color distributions that we would expect in
our survey if the galaxy populations are {\it non--evolving}. The
resulting distributions for the simulated cluster population are shown
in Fig.\ 3. No background contamination is included in this figure.
Table 4 lists the biweight location and scale estimators and their $1
\sigma$ uncertainties for these distributions.  The mean color as
measured by the biweight location estimator remains the same within
the measurement error as the distributions in Fig.\ 2; as expected,
the dispersion, as characterized by the biweight scale estimator,
increases by up to 15\%.

\subsection{Composite Color Distributions of the Palomar Clusters}

We now examine the color distributions of the 57 PDCS clusters.  For
each cluster candidate, we examine all of the galaxies within
$0.5~h^{-1}~{\rm Mpc}$ of the cluster center.  The composite CM
diagrams are shown in Fig.\ 4. We plot the aperture color
$(V_{4}-I_{4})_{ap}$ versus the aperture $I_{4}^{ap}$ magnitude.  We
have made composite CM diagrams separately for each of the five
fields.  The top three rows show the composites for each of the three
redshift intervals.  These plots include {\it both} cluster and
background galaxies. The bottom row of Fig.\ 4 shows the CM diagrams
for regions of the five fields which contain {\it no} detected
clusters (indicated as ``field''). Extinction corrections (see Paper
I) have been applied. In these diagrams, we have not corrected
$I_{4}^{ap}$ for the change in projected linear size of the
photometric aperture over each redshift interval because of the
redshift uncertainty. This correction corresponds to $\simless
0.30,~0.08$ and $0.03$ mag in the three respective intervals. The
correction due to color gradients in $B-V$ is small; this translates
into correction of $\simless 0.05$ mag in the $V_{4} - I_{4}$ color
(Sandage \& Visvanathan 1978; Silva \& Elston 1994).

In Fig.\ 5, we have combined the data of each PDCS field (Fig.\ 4) to
make composite color--magnitude diagrams for each of the three
redshift intervals. Even though we are contaminated by background
galaxies, it is clear that the CM diagram for the lowest redshift
interval (top row) shows a distinct locus of galaxies centered at
$(V_{4} - I_{4})_{ap} = 1.2 \pm 0.2$ and extending to magnitudes as
bright as $I_{4}^{ap} \sim 16$. This distinct feature is not clearly
observed in the higher redshift intervals (second and third rows). The
envelope in the lowest redshift interval corresponds well with the
color expected for a non--evolving early--type population.

Fig.\ 6 shows the corresponding color distributions for each of the
PDCS fields before (solid line histograms) and after (shaded
histograms) a statistical correction for background. Each cluster
color distribution was background subtracted using the background
color distribution determined from all regions of the field which
contain no detected clusters (bottom row of Fig.\ 6). The ``field''
regions of the $02^{\rm h}$ field contain slightly fewer galaxies
because a larger fraction of the field is excluded because either the
\v4 or \i4 band data does not exist.  We have co-added the 
background--corrected color distributions of each PDCS field (Fig.\ 6)
to make a global cluster composites for each redshift interval. Fig.\
7 shows the resulting distributions in each of the three redshift
intervals. The bar in each panel indicates the range of expected
(no--evolution) elliptical/S0 colors over this redshift interval (see
Fig.\ 1). Fig.\ 7 clearly reveals that we do not see the expected
effect of the large k--correction.

Since we cannot distinguish these galaxies morphologically, we use the
biweight location estimator as a representation of the mean galaxy
color and, presumably, the population of early--type galaxies. Table 5
lists the biweight location ($C_{BI}$) and scale estimator ($S_{BI}$)
for these distributions.  We have perturbed the expected field
contamination by $\pm 1\sigma$ and recomputed the distribution
characteristics. The resulting parameters are all within $1\sigma$ of
the original estimators in Table 5, indicating that we are not
adversely affected by the background subtraction.

As indicated in Fig.\ 7 and Table 5, there is a distinct peak at
$(V_{4} - I_{4})_{ap} \sim 1.3$ for clusters in the range $0.2 \le
z_{est} \le 0.4$. This color corresponds well to the expected
no--evolution color of present--day ellipticals at these redshifts
(Fig.\ 1). The distribution of colors in this redshift interval is
reasonably consistent with that expected from our simulations of the
no--evolution composite color distribution of a synthetic cluster
population (Fig.\ 3 and Table 3).  The results of the PDCS clusters
are also typical of other intermediate redshift clusters of galaxies
where a ridge line of early type galaxies, with colors similar to that
of present--day ellipticals, is obvious in the CM diagrams and color
distributions (BO; Couch \& Newell 1984; Millington \& Peach 1990;
Arag\'on-Salamanca et al.\ 1991; Luppino et al.\ 1991; Dressler \&
Gunn 1992; Le Borgne et al.\ 1992; Dressler et al.\ 1994; Molinari et
al.\ 1994; Smail et al.\ 1994).

Comparing our no--evolution predictions shown in Fig.\ 3 to the
resulting color distributions of the Palomar clusters shown in Fig.\
7, we see a distinct difference in the two highest redshift
intervals. In the interval $0.5 \le z_{est} \le 0.7$, the mean color
of the PDCS cluster galaxies is $(V_{4} - I_{4})_{ap} \sim 1.6$.  From
the no--evolution simulations of \S3.1, we would expect a clear peak
near $(V_{4} - I_{4})_{ap} \sim 2$ (Fig.\ 3).  The ``red envelope''
has apparently moved bluewards by $\sim 0.4$ mag. In the interval $0.8
\le z_{est} \le 1.2$, there are only a few ($\sim 5$) galaxies which
are as red as present--day ellipticals ($V_{4} - I_{4} \sim 3$ at
these redshifts).  The mean color, $(V_{4} - I_{4})_{ap} \sim 1.2$, is
bluer by $\sim 1.2$ mag than that predicted in the no--evolution
simulated cluster population (Table 4). If the mean color is an
accurate representation of the colors of the early--type cluster
population, this result indicates that these galaxies are
significantly bluer than the no--evolution predictions. This color
distribution may, however, be adversely influenced by two effects.
Firstly, late--type cluster members at high redshift would be
preferentially detected in the optical passbands.  This may mean that
the characteristic color is artificially bluer; however, this effect
is unlikely to account for the large color difference between our data
and the no--evolution predictions (compare Figs.\ 3 \& 7). Secondly,
as we only have estimated redshifts for most of the cluster sample,
there is the possibility that some of these high--redshift clusters
are actually superpositions of poor groups or clusters along the
line-of-sight.  Since there are only 5 clusters in the highest
redshift bin (Table 2), a misclassification of a few of these cluster
candidates would make the resulting color distribution significantly
bluer. Because of this possibility and the substantial threat of
contamination at these redshifts, we have used the Kolmogorov-Smirnov
(KS) test to confirm that the two high-redshift samples are not
consistent with being drawn from the field population. The probability
that the two color distributions are drawn from the field population
is less than 0.002\%.

In addition, we have explored a different approach in defining the
samples in the two highest redshift bins. We have imposed different
magnitude limits such that we are complete for the likely ${(V_{4} -
I_{4})}_{ap}$ colors of the early-type cluster population. These
limits are $I_{4}^{ap} = 20.5$ in the $0.5 \le z_{est} \le 0.7$
redshift interval and $I_{4}^{ap} = 20.0$ in the $0.8 \le z_{est} \le
1.2$ redshift interval. For a cluster at $z \sim 0.9$, the new
$I_{4}^{ap}$ magnitude limit should greatly improve the accurate
identification of the red cluster members (bottom left panel of Fig.\
2). From the new samples, we have created composite color
distributions in the same manner as described above. The biweight
location and scale estimators for the resulting color distributions
are $C_{BI} = 1.71 \pm 0.07$ and $S_{BI} = 0.51 \pm 0.05$ ($0.5 \le
z_{est} \le 0.7$) and $C_{BI} = 1.29 \pm 0.15$ and $S_{BI} = 0.70 \pm
0.10$ ($0.8 \le z_{est} \le 1.2$). The mean colors are reasonably
consistent within the measurement uncertainty with the results of the
original analysis; however, they are slightly redder as expected if
this technique improves our identification of the red cluster
galaxies. Still, the mean colors are bluer than the no--evolution
predictions of the synthetic cluster populations by $\sim 0.3$ mag at
$z \sim 0.6$ and $\sim 1.1$ mag at $z \sim 0.9$, respectively.

The estimated redshift distribution of the Palomar clusters is
examined in Paper I. This distribution is consistent with the
hypothesis that the typical, bright distant cluster galaxy is {\it
bluer} than a non--evolving elliptical at the cluster redshift. This
further supports the results of the color analysis presented here.

The observed blueing trend in the Palomar clusters has also been
observed by Rakos \& Schombert (1995) in the rest-frame Str\"omgren
$uvby$ filters and by A93 in the optical--infrared colors. A93
examined 10 rich clusters at $0.5 < z < 0.9$ and found that the red
envelope is bluer than present--day ellipticals by approximately 0.4,
0.5, and 1.3 mag in $V-K$ at the mean redshifts of $\langle z \rangle
= 0.56,~0.70,~{\rm and}~0.88$, respectively. Based on the simple
Bruzual \& Charlot (1996) evolutionary models presented in \S 3.3, the
observed $V-K$ color differences would imply a color shift (relative
to the no-evolution prediction) in the $V_{4} - I_{4}$ color of
roughly $0.4(\pm 0.2)$ and $1.0(\pm 0.2)$ at $z = 0.6$ and $0.9$,
respectively. These predictions are consistent within the photometric
and measurement errors with the observed color evolution in the
Palomar clusters.

\subsection{Comparison with Evolutionary Models}

Arag\'on-Salamanca et al.\ (1993), Dickinson (1995), and Rakos \&
Schombert (1995) have all noted that the observed evolutionary trend
in clusters at $z \simgreat 0.4$ is consistent with simple passive
evolution of an old stellar population which was originally formed in
a single burst of star-formation at an epoch of $z \simgreat 2$ (see
also Bower et al.\ 1992a,b; Charlot \& Silk 1994; Pehre et al.\ 1995;
Ellis et al.\ 1996). We examine these conclusions by comparing the
results of our color analysis to the population synthesis models of
Bruzual \& Charlot (e.g.\ Bruzual 1983; Bruzual \& Charlot 1993, 1996
and references therein).  The free parameters in these models are the
initial mass function (IMF) and the star formation rate (SFR). We
choose the traditional Scalo (1986) IMF with lower and upper mass
limits of 0.1 and 125 $M_{\odot}$, respectively. For the SFR, we
explore two possible rates : (1) a burst of constant star formation
for a initial period $\tau$ ($c$--model; Bruzual 1983); and (2) an
exponentially decaying SFR such that a fraction $\mu = 1 - e^{-1 {\rm
Gyr}/\tau}$ of the galaxy mass is converted into stars after the first
Gyr ($\mu$--model; Bruzual 1983). The evolutionary calculations have
been made using the Bruzual \& Charlot (1996) synthesis code. The age
of the galaxy is determined by the redshift ($z_{f}$) at which the
epoch of star formation began. We relate look-back time to redshift
using $H_{o} = 75~{\rm km~s^{-1}~Mpc^{-1}}$ and $q_{o} = \{0.0~0.5\}$.

In Fig.\ 8, we show the results for two models with (1) $\tau = 1~{\rm
Gyr}$ and (2) $\mu = 0.5$ for three formation epochs of $z_{f} =
\{1~2~10\}$. Given the large redshift uncertainty, the color evolution
in the Palomar clusters is consistent with passive evolution of a
single burst of star formation at $z_{f} \approx 2$. Both models
provide a reasonable fit to the color evolution at $z \simless 0.6$,
implying formation epochs of $z_{f} \simgreat 2$. Previous
observations of distant clusters in the optical--infrared colors
indicate similar formation epochs (A93; Dickinson 1995; Rakos \&
Schombert 1995).

The $\mu$--model provides a better fit to the data in the highest
redshift bin as more evolution is predicted at $z \sim 0.9$.  As noted
in \S3.2, since optical passbands at high-redshift sample the
rest-frame UV light which is most affected by recent bursts of star
formation, the color distribution in the highest redshift bin will be
biased toward late-type cluster galaxies. Therefore, the
characteristic (or mean) color of this distribution may not accurately
represent the early-type cluster population but may actually be {\it
bluer}. This would, of course, affect how well a particular model fits
the optical colors. In light of this, we have examined a combination
of evolutionary models representing a mix of early and late--type
galaxies, i.e.\ the $\tau = 1~{\rm Gyr}$ model for the ellipticals and
a constant SFR model for the spirals (see Bruzual \& Charlot
1996). Mixed models with a cluster population of $\simgreat 50\%$
spirals match well the color in the highest redshift bin.  These
models are also consistent with formation epochs of $z_{f} \simgreat
2$ for the early--type population.  Observations in the infrared bands
are more indicative of a long-lived stellar population and are thereby
more suited to examining the red, ``passive'' early-type population. A
deep optical--infrared survey of the Palomar clusters at $z \simgreat
0.6$ is presently underway (Lubin et al.\ 1996).

Regardless of the specific forms of the evolutionary models, we have
shown in \S3.2 that our data are inconsistent with a non-evolving
population of early-type galaxies. In comparison with the simple
passive evolution models presented here, the observed evolution at $z
\simless 1$ suggests that these galaxies formed at redshifts of $z
\simgreat 2$, consistent with previous observations of distant
clusters of galaxies. This result is supported by direct observations
of very high--redshift ($z > 3$) star-forming galaxies (Steidel et
al.\ 1996).

\subsection{Composite Radial Color Distributions}

We would also like to examine the radial dependence on the cluster
color distribution. In order to improve our cluster signal-to-noise at
large radii from the cluster center, we examine an even richer subset
of PDCS clusters; that is, we include only those clusters with a
filter richness measure $\Lambda_{cl} \ge 70$ in either of the \v4 or
\i4 bands (see \S 2.2). This corresponds to a sample of 10, 14, and 5
clusters in each of the three redshift intervals.  In Fig.\ 9, we
present the background-corrected composite color distributions as a
function of three radial zones : (1) $r \le 0.25$, (2) $0.25 < r \le
0.50$, and (3) $0.50 < r \le 0.75~h^{-1}~{\rm Mpc}$.  Table 6 lists
the biweight location and scale estimators for these
distributions. There is no obvious trend in the data, except perhaps
in the intermediate redshift interval of $0.5 \le z_{est} \le
0.7$. Here the mean color, as defined by the biweight location
estimator, gets slightly bluer as a function of radial position. A
strong blueing trend with distance from the cluster center has been
previously observed by BO in intermediate redshift ($0.35 < z < 0.5$)
clusters. In addition, one might expect the dispersion, as
characterized by the biweight scale estimator, to increase with radius
as the tight color--magnitude relation for early--type galaxies is
characteristic of the cluster cores. This is not clearly observed in
the PDCS clusters. We note that at large radii the variance may
increase and the color may become artificially bluer if we have not
accurately removed the contribution of the mostly late--type
background population.

\section{The Butcher--Oemler Effect}

Butcher \& Oemler (1984) have systematically studied 33 clusters of
galaxies with redshifts between 0.003 and 0.54 to examine the
evolution of the colors of the cluster populations. They found that
the ``fraction of blue galaxies'' ($f_{b}$) increases with redshift
for both open and compact clusters. BO defined $f_{b}$ as follows;
they examined only those galaxies which are brighter than $M_{V} =
-20$ and within a circular area containing the inner 30\% of the total
cluster population (as characterized by the radius $R_{30}$ where
$R_{n}$ is the radius containing n\% of the cluster's projected galaxy
distribution within $\sim 0.75~h^{-1}~{\rm Mpc}$). $f_{b}$ is then
defined as the fraction of galaxies whose {\it rest--frame} $B-V$
colors are at least 0.2 mag bluer than the ridge line of the
early--type galaxies at that magnitude. The blue fraction is also
dependent on the degree of central concentration (Oemler 1974;
Dressler 1980; BO). BO measure central concentration by the
compactness parameter $C \equiv {\rm log}~(R_{60}/R_{20})$ [see sample
profiles in Butcher \& Oemler 1978]. $C \simgreat 0.4$ corresponds
roughly to a cluster with a centrally concentrated galaxy
population. At the present epoch, compact clusters contain few spirals
in their cores ($f_{b} \sim 0.03 \pm 0.01$ for clusters with $z
\simless 0.1$).  In compact clusters at $z \sim 0.5$, the blue
fraction has increased to $\sim 0.25$.  Clusters whose density
distributions approximate those of a uniform--density sphere have $C
\approx 0.3$. At the present epoch, these open clusters have large
spiral populations ($f_{b} \sim 0.2$). By $z \sim 0.4$, open clusters
have $f_{b} \sim 0.35$ (Oemler 1974; Butcher \& Oemler 1978; Dressler
1980; BO).

We perform an identical study on the 57 PDCS cluster candidates.  In
order to compare as closely as possible with the original BO analysis,
we need to transform our photometric observations to the cluster rest
frame at each of the estimated redshifts ($z_{est}$).  We have
transferred to the cluster $V$ rest frame by using the observed \i4
magnitudes. The absolute magnitude in the \i4 band is given by

\begin{equation}
M_{I_{4}} = I_{4} - 5~{\rm log}~d_{L} - 25 - K_{I_{4}}
\end{equation}

\noindent where $d_{L}$ is the luminosity distance in $h^{-1}~{\rm Mpc}$,
and $K_{I_{4}}$ is the k--correction in the \i4 band (see Fig.\ 1).
We can write the identity $M_{V} - M_{I_{4}} = {(V - I_{4})}_{o}$,
where ${(V - I_{4})}_{o}$ is the rest frame color. From this relation
and Eq.\ (5), the absolute $V$ magnitude is then given by

\begin{equation}
M_{V} = I_{4} - 5~{\rm log}~d_{L} - 25 - K_{I_{4}} + {(V-I_{4})}_{o}\nonumber
\end{equation}

At each of the estimated redshifts of our cluster candidates (Table
1), the k--correction (Fig.\ 1) and rest frame ${(V - I_{4})}_{o}$
color are computed by convolving the the elliptical/S0 SED of Coleman,
Wu \& Weedman (1980) with the system filter bandpasses. The \v4 and
\i4 bandpasses are shown in Fig.\ 1 of Paper I, and the $B$ and $V$
filters are given in Landolt (1992).  From Eq.\ (6), we can determine
the \i4 magnitude which corresponds to the BO absolute magnitude limit
of $M_{V} = -20$. At $z_{est} \approx 0.6$, the BO magnitude limit is
{\it fainter} than the PDCS survey limit.  Though we still present our
calculations of $f_{b}$ for clusters at $z_{est} \simgreat 0.6$, we
are not complete at these redshifts. Finally, using the SEDs of
Coleman, Wu \& Weedman (1980) and the appropriate band passes, we
determine the color difference $\Delta(V_{4} - I_{4})$ which
corresponds to a color difference of $\Delta(B-V) = 0.2$ at each of
the estimated redshifts.

For each candidate cluster, we examine those galaxies brighter than
$M_{V} = -20$. In order to determine the compactness parameter $C$ and
the fiducial radius $R_{30}$, we use the cluster profiles presented in
Paper II. That is, we have binned the galaxies out a radius of
$1.0~h^{-1}~{\rm Mpc}$ and fit the resulting background-subtracted
cluster profiles to a King model (see Paper II for details). From the
best--fit King model, we analytically calculate $R_{30}$ and the
compactness parameter $C$ (as defined above). Table 1 lists the
compactness parameter for the cluster candidates.

The resulting background-subtracted \v4 -- \i4 color distributions are
examined. Following the BO analysis, we determine the ridge line of
the early--type population for each cluster. We take the {\it median}
color of each distribution as this characteristic color. The fraction
of blue galaxies ($f_{b}$) for each cluster is then calculated
relative to this color by using the corresponding $\Delta(V_{4} - I_{4})$
analytically determined above.  The resulting $f_{b}$ of each cluster
is listed in Table 1. Fig.\ 10 shows $f_{b}$ versus the estimated
redshift ($z_{est}$). We indicate with different points a richer
subset of clusters ($\Lambda_{cl} \ge 70$ in either the \v4 or \i4
band). The median values of the blue fraction (indicated by large
boxed crosses in Fig.\ 10) are $f_{b} = \{0.18~ 0.29\}$ in the $0.2 \le
z_{est} \le 0.4$ and the $0.5 \le z_{est} \le 0.7$ redshift intervals,
respectively.

In order to estimate the error in $f_{b}$, we have calculated the
standard error assuming Poissonian statistics; in addition, we have
perturbed the background counts by $\pm 1\sigma$ and recalculated
$f_{b}$.  The uncertainty in $f_{b}$ represents the {\it range} of
observed values for a particular cluster. We have not corrected the
color distributions for the color--magnitude effect in E/S0
galaxies. This effect is the slight blueing of E/S0 galaxy colors as
the galaxy absolute magnitude becomes fainter (Sandage \& Visvanathan
1978). It broadens the color distribution of the early--type
population and might, therefore, cause an artificial increase in
$f_{b}$.  We have tested this effect by recalculating $f_{b}$ with a
color difference $\Delta(B-V) = 0.4$. The resulting $f_{b}$ values are
all within the lower error bars.

The scatter in $f_{b}$ is large, as is the uncertainty in the
estimated redshift ($\sigma_{z_{est}} \simless 0.2$); however, if we
take the data points strictly at face value, there is a correlation
between the blue fraction $f_{b}$ and the estimated redshift
$z_{est}$.  We have quantified this by calculating the Spearman
rank-order correlation coefficient for our data points. The
correlation is significant at a 99.98\% confidence level.  In order to
include the parameter uncertainties in this calculation, we have
created 200 new datasets by randomly adding the appropriate error to
each data point, assuming that the uncertainties in both parameters
are gaussian. A correlation coefficient and significance level are
then calculated for each new sample. The average significance of the
correlation between $z_{est}$ and $f_{b}$ is 96.2\% ($\sim 2 \sigma$).
The purpose of this statistical analysis is to show that there is a
{\it monotonic increase} in the blue fraction with estimated redshift,
not necessarily that there is a linear correlation between the two
parameters as we have no reason to believe that the relationship is
linear.

Our values of $f_{b}$ as a function of redshift are consistent with
previous observations of intermediate and high--redshift clusters. We
find blue fractions of $f_{b} \sim 0.05 - 0.2$ in clusters at $0.2
\simless z \simless 0.4$, consistent with the results of BO. By $z
\sim 0.5$, $f_{b}$ reaches $\sim 0.3$. The trend in Fig.\ 10 suggests
that, at redshifts of $z \simgreat 0.6$, the fraction of blue galaxies
may be greater than 0.4. Rakos \& Schombert (1995) find that the
fraction of blue galaxies, as defined in the rest--frame Str\"omgren
colors, is $f_{b} \sim 0.4 - 0.6$ at $z \sim 0.6$ and increases to
values as high as 0.8 in clusters at $z = 0.9$.

BO found that the fraction of blue galaxies in open clusters increases
with redshift at a rate similar to that in compact clusters, though
$f_{b}$ was systematically larger (see \S 1). We have examined this
relation in our data. Fig.\ 11 shows $f_{b}$ as a function of
estimated redshift. We indicate with different symbols three ranges of
the compactness parameter $C$ (defined above).  Filled, dotted, and
open circles indicate compact clusters ($C \ge 0.4$), intermediate
clusters ($0.35 \le C < 0.4$), and open clusters ($C < 0.35$),
respectively, as specified by BO. The typical PDCS clusters is compact
(though not azimuthally symmetric; see Paper II). Considering the
large uncertainty, there is no clear indication that the open or
intermediate clusters have systematically higher blue fractions.

\section{Summary}

We have examined the color evolution of the galaxy populations of the
richest candidate clusters of galaxies from the Palomar Distant
Cluster Survey. The sample contains 57 intermediate and high--redshift
clusters of galaxies, the largest statistical sample of distant
clusters. Our principal conclusions are summarized below.

\newcounter{discnt}

\begin{list}
{\arabic{discnt}.}  {\usecounter{discnt}}

\item For intermediate redshift ($z \simless 0.4$) clusters, we find
that there is a distinct ridge line or ``red envelope'' in the
color--magnitude diagram. The color of this locus corresponds well
with the expected no--evolution color of present--day ellipticals at
these redshifts. This result is consistent with previous optical
studies of intermediate redshift clusters.

\item At progressively higher redshifts, this red envelope, as
characterized by the mean galaxy color, has shifted bluewards compared
to the no--evolution prediction. By $z \sim 0.9$, there are only a few
cluster galaxies which are as red in their rest--frame as present-day
ellipticals.  Through simple Monte--Carlo simulations of a synthetic
non--evolving cluster population, we have confirmed that we are able
to observe these very red members if they are present at these
redshifts.  The blueing trend of the mean galaxy color by $\sim 0.4$
mag at $z \sim 0.6$ and over 1 mag at $z \sim 0.8$ is consistent with
that previously observed in the optical--infrared colors.  A
comparison between the observed color evolution in the Palomar
clusters with simple Bruzual \& Charlot (1996) models of passive aging
of an old stellar population formed in a single burst of star
formation indicates formation epochs of $z \simgreat 2$ for the
early-type cluster galaxies.

\item Though the uncertainties are large, the Butcher--Oemler effect
is observed in the Palomar clusters. We find that the fraction of blue
galaxies increases with the estimated redshift of the cluster at a
96.2\% ($2 \sigma$) confidence level. We observe blue fractions of
$f_{b} \sim 0.05 - 0.2$ in clusters at redshifts of $0.2 \simless z
\simless 0.4$. These fractions are consistent with that found
previously by Butcher \& Oemler (1984). Our calculations of $f_{b}$
are only complete to redshifts of $z \sim 0.6$; at these redshifts,
$f_{b} \sim 0.3 - 0.4$. The observed trend indicates that clusters at
$z \simgreat 0.6$ can have blue fractions which are greater than
0.4. This result is consistent with that found by Rakos \& Schombert
(1995). They have examined 17 clusters over a redshift range similar
to ours and find that the blue fraction $f_{b}$ increases to 0.8 in
clusters at $z \sim 0.9$.

\end{list}

\vskip 0.3cm

The referee Jim Schombert is graciously thanked for his insightful
comments and an extremely productive visit. It is also a pleasure to
thank Marc Postman for his continual guidance and essential scientific
contributions to this paper and Ian Smail for his special attention
and thorough review of this text. Neta Bahcall, Jim Gunn, Michael
Strauss, and Ed Turner are thanked for providing invaluable comments
on a preliminary version. The following people are acknowledged for
their generous contributions : Robert Lupton for his useful comments
and aid in photometric conversions, Don Schneider for providing the
spectral energy distributions and response functions, St\'ephane
Charlot and Ann Zabludoff for supplying and explaining the Bruzual \&
Charlot synthesis codes, and St\'ephane Courteau for being available
for scientific consultations. LML graciously acknowledges support from
a Carnegie Fellowship and NASA contract NGT--51295.

\vfill
\eject

\vfill
\eject

\centerline{\bf Figure Captions}
 
\newcounter{discnt2}
 
\begin{list}
{{\bf Figure \arabic{discnt2}} :}  {\usecounter{discnt2}}

\item The no-evolution $V_{4} - I_{4}$ color and the \v4 and \i4 band
k-corrections as a function of redshift and morphological type as
calculated using the spectral energy distributions from Coleman, Wu \&
Weedman (1980).

\item Left : the expected color--magnitude relation for a simulated
no--evolution cluster population as a function of redshift (see \S
3.1). The cluster redshift is indicated in the upper left corner of
each panel. E/S0s are indicated by filled circles. Sab, Sbc, Scd, and
Irr galaxies are indicated by circles, squares, triangles, and stars,
respectively. The magnitude limits of the PDCS ($V_{4}^{ap} \approx
22.8$ and $I_{4}^{ap} \approx 21.4$) are indicated by solid
lines. Right : the resulting color distributions. Solid line
histograms represent the total color distribution, while shaded
histograms represent the color distributions that would be observed in
our survey.

\item The expected composite color distributions in each of the three
redshift intervals for the simulated no--evolution cluster populations
(see \S 3.1).  Solid line histograms represent the total color
distribution, while shaded histograms represent the color distribution
that would be observed in our survey. For clarity, the {\it observed}
color distribution in the highest redshift interval is expanded in the
small window in the bottom panel.

\item Composite color--magnitude diagrams for each of the five PDCS
fields. The aperture color, $(V_{4} - I_{4})_{ap}$, is plotted against
the aperture \i4 magnitude, $I_{4}^{ap}$. The aperture magnitude
limits have been applied. The top three panels show the composite CM
diagrams for each of the three redshift intervals (see \S3.2).  Table 2
lists the number of cluster candidates in each panel. The bottom
panels show the CM diagrams of regions of the five fields which
contain no cluster galaxies (indicated as ``field'').

\item Composite color--magnitude diagrams of the Palomar clusters in
each of the three redshift intervals. The aperture color, $(V_{4} -
I_{4})_{ap}$, is plotted against the aperture \i4 magnitude,
$I_{4}^{ap}$. The aperture magnitude limits have been applied.  The
bottom panel shows the CM diagram of sample regions of the five
fields which contain no cluster galaxies (indicated as ``field'').

\item Composite aperture color distributions for each of the five
PDCS fields. The cluster galaxy color distributions are shown before
(solid line histograms) and after (shaded histograms) a statistical
correction for the background. The aperture magnitude limits have been
applied. The top three panels show the composite CM diagrams for each
of the three redshift intervals (see \S3.2).  The bottom panels show
the CM diagrams of regions of the five fields which contain no cluster
galaxies (indicated as ``field'').

\item Composite aperture color distributions of the Palomar clusters
in each of the three redshift intervals.  Background has been
subtracted. The redshift interval is listed in the upper left of each
panel. The expected range of {\it no--evolution} E/S0s colors (Fig.\
2) for each redshift interval is indicated by a bar (see \S 3.2).

\item Comparison of the characteristic colors of the Palomar clusters
for each of the three redshift bins (see \S 3) with the results of the
Bruzual \& Charlot models (see \S 3.3) : a $\tau = 1~{\rm Gyr}$ burst
of star formation (upper panel) and an exponentially decaying star
formation rate with $\mu = 0.5$ (lower panel). The lines represent
different epochs of the initial star formation : $z_{f} = 1$ (solid
lines), $z_{f} = 2$ (dotted lines), and $z_{f} = 10$ (dashed lines)
with $q_{o} = 0$ (thin lines) and $q_{o} = 0.5$ (thick lines). The
vertical errors indicate the $1 \sigma$ confidence limits on the
characteristic color of the background-subtracted galaxy distribution
(see \S 3.1). The horizontal errors indicate the approximate range of
estimated redshifts in each bin.

\item Composite aperture color distributions versus radial distance
for a richer ($\Lambda_{cl} \ge 70$) sample of the PDCS clusters (see
\S 3.4).  Columns indicate each of the three redshift intervals.  The
radial range (in $h^{-1}~{\rm Mpc}$) around the cluster center is
indicated in the upper left-hand corner of each panel.

\item The fraction of blue galaxies ($f_{b}$) versus estimated
redshift ($z_{est}$) for the 57 cluster candidates. In order to show
all points, we have randomly offset the cluster $z_{est}$ by less than
$\pm 0.05$.  A richer subsample of clusters ($\Lambda_{cl} \ge 70$) is
indicated by the filled circles. $f_{b}$ is only complete to $z_{est}
\approx 0.6$. The errors in $f_{b}$ represent the range of possible
values. The estimated redshift uncertainty is indicated in the upper
right hand corner. The median values in the first two redshift bins
are designated by the large boxed crosses. $f_{b}$ and $z_{est}$ are
correlated at a 96.2\% ($\sim 2 \sigma$) confidence level (see \S 4).

\item The fraction of blue galaxies ($f_{b}$) versus estimated
redshift for the 57 cluster candidates. The symbols indicate three
compactness ($C$) ranges of the cluster candidates (see \S 4). Filled
circles, compact clusters ($C \ge 0.4$); open circles, open clusters
($C < 0.35$); dotted circles, intermediate clusters ($0.35 \le C <
0.4$).

\end{list}

\vfill
\eject

\begin{table}
\begin{center}
Table 1 : Properties of the Sample of PDCS Clusters
\end{center}
\begin{center}
\begin{tabular}[t]{ccrccc}
\hline
\hline
Cluster	&Redshift	& \multicolumn{2}{c}{Richness} 	& Compactness	& Blue Fraction \\
ID \#	&$z_{est}$	& \multicolumn{2}{c}{$\Lambda_{cl}$}& $C$		&$f_{b}$	\\
\hline
001     &   0.6  & 61.5   & &    $0.60^{+0.09}_{-0.09}$      &$0.28^{+0.26}_{-0.12}$ \\
002     &   0.4  & 52.1   & &    $0.48^{+0.05}_{-0.05}$      &$0.27^{+0.09}_{-0.18}$ \\
003     &   0.6  & 88.5   & &    $0.32^{+0.09}_{-0.02}$      &$0.15^{+0.11}_{-0.12}$ \\
004     &   0.6  & 94.9   & &    $0.35^{+0.03}_{-0.02}$      &$0.33^{+0.06}_{-0.12}$ \\
006     &   0.5  & 62.9   & &    $0.33^{+0.02}_{-0.02}$      &$0.35^{+0.05}_{-0.09}$ \\
008     &   0.6  & 75.3   & &    $0.38^{+0.13}_{-0.11}$      &$0.25^{+0.06}_{-0.09}$ \\
009     &   0.4  & 58.1   & &    $0.40^{+0.02}_{-0.02}$      &$0.20^{+0.12}_{-0.20}$ \\
010     &   0.3  & 58.8   & &    $0.38^{+0.15}_{-0.11}$      &$0.27^{+0.10}_{-0.18}$ \\
011     &   0.4  & 104.5  & &    $0.48^{+0.08}_{-0.09}$      &$0.18^{+0.12}_{-0.09}$ \\
012     &   0.3  & 75.5   & &    $0.36^{+0.04}_{-0.06}$      &$0.09^{+0.06}_{-0.09}$ \\
014     &   0.4  & 46.4   & &    $0.49^{+0.04}_{-0.06}$      &$0.18^{+0.06}_{-0.05}$ \\
015     &   1.1  & 78.5   & &    $0.52^{+0.04}_{-0.05}$      &$0.37^{+0.10}_{-0.07}$ \\
016     &   0.5  & 108.6  & &    $0.57^{+0.02}_{-0.02}$      &$0.18^{+0.06}_{-0.15}$ \\
017     &   0.7  & 115.0  & &    $0.45^{+0.03}_{-0.02}$      &$0.21^{+0.07}_{-0.12}$ \\
018     &   0.4  & 54.0   & &    $0.52^{+0.02}_{-0.02}$      &$0.15^{+0.08}_{-0.08}$ \\
019     &   0.6  & 66.3   & &    $0.75^{+0.05}_{-0.23}$      &$0.22^{+0.14}_{-0.04}$ \\
020     &   0.4  & 50.2   & &    $0.25^{+0.04}_{-0.04}$      &$0.28^{+0.19}_{-0.06}$ \\
021     &   0.3  & 45.2   & &    $0.58^{+0.02}_{-0.02}$      &$0.20^{+0.05}_{-0.20}$ \\
022     &   0.4  & 64.7   & &    $0.34^{+0.02}_{-0.02}$      &$0.27^{+0.05}_{-0.05}$ \\
023     &   0.2  & 44.6   & &    $0.44^{+0.17}_{-0.14}$      &$0.21^{+0.07}_{-0.07}$ \\
024     &   0.4  & 66.8   & &    $0.32^{+0.02}_{-0.05}$      &$0.11^{+0.05}_{-0.07}$ \\
025     &   0.3  & 49.5   & &    $0.42^{+0.03}_{-0.02}$      &$0.10^{+0.09}_{-0.10}$ \\
030     &   0.3  & 46.5   & &    $0.27^{+0.03}_{-0.02}$      &$0.13^{+0.05}_{-0.05}$ \\
031     &   1.1  & 120.0  & &    $0.37^{+0.04}_{-0.07}$      &$0.33^{+0.10}_{-0.23}$ \\
033     &   0.5  & 42.5   & &    $0.60^{+0.04}_{-0.18}$      &$0.26^{+0.15}_{-0.20}$ \\
034     &   0.3  & 62.0   & &    $0.49^{+0.04}_{-0.04}$      &$0.11^{+0.03}_{-0.05}$ \\
035     &   0.6  & 67.5   & &    $0.46^{+0.02}_{-0.02}$      &$0.40^{+0.30}_{-0.20}$ \\
036     &   0.3  & 54.7   & &    $0.61^{+0.05}_{-0.11}$      &$0.09^{+0.05}_{-0.10}$ \\
037     &   0.6  & 60.0   & &    $0.70^{+0.06}_{-0.17}$      &$0.25^{+0.08}_{-0.12}$ \\
\hline								
\end{tabular}							
\end{center}							
\end{table}

\begin{table}
\begin{center}
Table 1 -- Continued
\end{center}
\begin{center}
\begin{tabular}[t]{ccrlcc}
\hline
\hline
Cluster	&Redshift	&\multicolumn{2}{c}{Richness}	& Compactness	& Blue Fraction \\
ID \#	&$z_{est}$	&\multicolumn{2}{c}{$\Lambda_{c}$}& $C$		&$f_{b}$	\\
\hline
038     & 0.3  & 47.7   &&  $0.61^{+0.02}_{-0.02}$       &$0.33^{+0.05}_{-0.22}$   \\
039     & 0.6  & 63.3   &&  $0.72^{+0.04}_{-0.03}$       &$0.09^{+0.11}_{-0.09}$   \\
041     & 0.6  & 59.6   &&  $0.63^{+0.07}_{-0.07}$       &$0.36^{+0.13}_{-0.18}$   \\
042     & 0.6  & 77.4   &&  $0.61^{+0.27}_{-0.19}$       &$0.31^{+0.20}_{-0.20}$   \\
047     & 0.3  & 38.5   &&  $ --		 $       &$--		     $   \\
048     & 0.3  & 28.4   &&  $0.47^{+0.04}_{-0.05}$       &$0.20^{+0.07}_{-0.15}$   \\
049     & 0.2  & 51.2   &&  $ --		 $       &$--		     $   \\
050     & 0.5  & 96.8   &&  $0.46^{+0.08}_{-0.06}$       &$0.29^{+0.07}_{-0.05}$   \\
051     & 0.4  & 76.5   &&  $0.51^{+0.05}_{-0.06}$       &$0.17^{+0.05}_{-0.05}$   \\
052     & 0.5  & 114.3  &&  $0.71^{+0.02}_{-0.03}$       &$0.29^{+0.06}_{-0.20}$   \\
053     & 0.2  & 19.4   &&  $0.32^{+0.02}_{-0.02}$       &$0.00^{+0.05}_{-0.05}$   \\
054     & 0.3  & 70.5   &&  $0.51^{+0.03}_{-0.03}$       &$0.06^{+0.07}_{-0.06}$   \\
055     & 0.3  & 72.2   &&  $0.37^{+0.02}_{-0.02}$       &$0.18^{+0.05}_{-0.07}$   \\
056     & 0.5  & 61.4   &&  $0.47^{+0.04}_{-0.06}$       &$0.00^{+0.20}_{-0.05}$   \\
057     & 0.8  & 72.7   &&  $0.64^{+0.06}_{-0.02}$       &$0.22^{+0.05}_{-0.09}$   \\
059     & 0.8  & 115.3  &&  $0.57^{+0.14}_{-0.18}$       &$0.13^{+0.15}_{-0.15}$   \\
061     & 0.3  & 47.7   &&  $0.57^{+0.04}_{-0.06}$       &$0.19^{+0.11}_{-0.11}$   \\
062     & 0.4  & 80.3   &&  $0.32^{+0.02}_{-0.02}$       &$0.13^{+0.05}_{-0.05}$   \\
063     & 0.6  & 103.7  &&  $0.32^{+0.02}_{-0.02}$       &$0.29^{+0.15}_{-0.15}$   \\
065     & 0.4  & 61.0   &&  $0.43^{+0.22}_{-0.11}$       &$0.15^{+0.07}_{-0.07}$   \\
066     & 0.4  & 34.5   &&  $0.56^{+0.14}_{-0.15}$       &$0.27^{+0.20}_{-0.15}$   \\
067     & 0.5  & 45.7   &&  $0.40^{+0.04}_{-0.10}$       &$0.57^{+0.11}_{-0.29}$   \\
068     & 0.5  & 59.4   &&  $0.40^{+0.02}_{-0.02}$       &$0.07^{+0.13}_{-0.07}$   \\
069     & 0.3  & 43.5   &&  $0.42^{+0.02}_{-0.05}$       &$0.00^{+0.09}_{-0.09}$   \\
071     & 0.7  & 95.5   &&  $0.32^{+0.14}_{-0.05}$       &$0.33^{+0.15}_{-0.06}$   \\
072     & 0.6  & 85.1   &&  $0.40^{+0.07}_{-0.08}$       &$0.42^{+0.14}_{-0.16}$   \\
075     & 0.9  & 147.7  &&  $0.54^{+0.03}_{-0.03}$       &$0.21^{+0.12}_{-0.06}$   \\
076     & 0.3  & 54.0   &&  $0.43^{+0.02}_{-0.02}$       &$0.33^{+0.10}_{-0.10}$   \\
\hline								
\end{tabular}							
\end{center}							
\end{table}

\begin{table}
\begin{center}
Table 2 : The Number of PDCS Clusters in Each Field
\end{center}
\begin{center}
\begin{tabular}[t]{lcccccc}
\hline
\hline
Redshift Interval &$00^{{\rm h}}$&$02^{{\rm h}}$&$09^{{\rm h}}$&$13^{{\rm h}}$&$16^{{\rm h}}$&total \\
\hline
$0.2 \le z_{est} \le 0.4$       &5	&8	&4	&8	&4	&29\\
$0.5 \le z_{est} \le 0.7$       &5	&3	&6	&5	&4	&23\\
$0.8 \le z_{est} \le 1.2$       &0	&1	&1	&2	&1	&5\\
\hline
\end{tabular}
\end{center}
\end{table}

\begin{table}
\begin{center}
Table 3 : Statistical Descriptors of the $V_{4} - I_{4}$ Color
Distribution of a Simulated Cluster Population
\end{center}
\begin{center}
\begin{tabular}[t]{ccccc}
\hline
\hline
$z$ 	& \multicolumn{2}{c}{Total} & \multicolumn{2}{c}{Observed} \\
	& $C_{BI}$& $S_{BI}$ & $C_{BI}$& $S_{BI}$ \\
\hline
0.3	&$1.19 \pm 0.05$&$0.44 \pm 0.03$&$1.20 \pm 0.05$& $0.44 \pm 0.03$ \\
0.6	&$2.10 \pm 0.07$&$0.56 \pm 0.04$&$2.20 \pm 0.11$& $0.54 \pm 0.07$ \\
0.9	&$2.60 \pm 0.09$&$0.65 \pm 0.06$&$2.40 \pm 0.26$& $0.77 \pm 0.16$ \\
\hline
\end{tabular}
\end{center}
\end{table}

\begin{table}
\begin{center}
Table 4 : Statistical Descriptors of the Composite $V_{4} - I_{4}$
Color Distributions of Simulated Cluster Populations
\end{center}
\begin{center}
\begin{tabular}[t]{ccccc}
\hline
\hline
Redshift Interval & \multicolumn{2}{c}{Total} & \multicolumn{2}{c}{Observed} \\
	& $C_{BI}$& $S_{BI}$ & $C_{BI}$& $S_{BI}$ \\
\hline
$0.2 \le z \le 0.4$&$1.19 \pm 0.01$&$0.50 \pm 0.02$&$1.20 \pm 0.02$& $0.50 \pm 0.02$\\
$0.5 \le z \le 0.7$&$2.04 \pm 0.02$&$0.60 \pm 0.03$&$1.99 \pm 0.03$& $0.59 \pm 0.02$\\
$0.8 \le z \le 1.2$&$2.58 \pm 0.04$&$0.64 \pm 0.03$&$2.44 \pm 0.10$& $0.80 \pm 0.05$\\
\hline
\end{tabular}
\end{center}
\end{table}

\begin{table}
\begin{center}
Table 5 : Statistical Descriptors of the Composite $V_{4} - I_{4}$
Color Distributions of the PDCS Clusters
\end{center}
\begin{center}
\begin{tabular}[t]{ccc}
\hline
\hline
Redshift Interval		& $C_{BI}$& $S_{BI}$ \\
\hline
$0.2 \le z_{est} \le 0.4$       &$1.34 \pm 0.02$& $0.45 \pm 0.05$\\
$0.5 \le z_{est} \le 0.7$       &$1.54 \pm 0.06$& $0.54 \pm 0.05$\\
$0.8 \le z_{est} \le 1.2$	&$1.21 \pm 0.07$& $0.50 \pm 0.09$\\
\hline
\end{tabular}
\end{center}
\end{table}

\begin{table}
\begin{center}
Table 6 : Statistical Descriptors of the Composite $V_{4} - I_{4}$
Color Distributions of the PDCS Clusters as a Function of
Radius
\end{center}
\begin{center}
\begin{tabular}[t]{ccccccc}
\hline
\hline
Radial Interval& \multicolumn{2}{c}{$0.2 \le z_{est} \le 0.4$} &\multicolumn{2}{c}{$0.5 \le z_{est} \le 0.7$} & \multicolumn{2}{c}{$0.8 \le z_{est} \le 1.2$}  \\
$(h^{-1}~{\rm Mpc})$& $C_{BI}$& $S_{BI}$ & $C_{BI}$& $S_{BI}$ & $C_{BI}$& $S_{BI}$ \\
\hline
$r \le 0.25$		&$1.34 \pm 0.04$& $0.44 \pm 0.03$ &$1.60 \pm 0.03$& $0.52 \pm 0.04$ &$1.27 \pm 0.08$& $0.48 \pm 0.08$ \\
$0.25 \le r \le 0.50$   &$1.32 \pm 0.05$& $0.45 \pm 0.05$ &$1.46 \pm 0.04$& $0.55 \pm 0.04$ &$1.18 \pm 0.07$& $0.52 \pm 0.09$ \\
$0.50 \le r \le 0.75$   &$1.41 \pm 0.03$& $0.37 \pm 0.06$ &$1.36 \pm 0.04$& $0.48 \pm 0.04$ &$1.31 \pm 0.09$& $0.50 \pm 0.07$ \\
\hline
\end{tabular}
\end{center}
\end{table}


\begin{figure}
\centerline{
\epsfbox{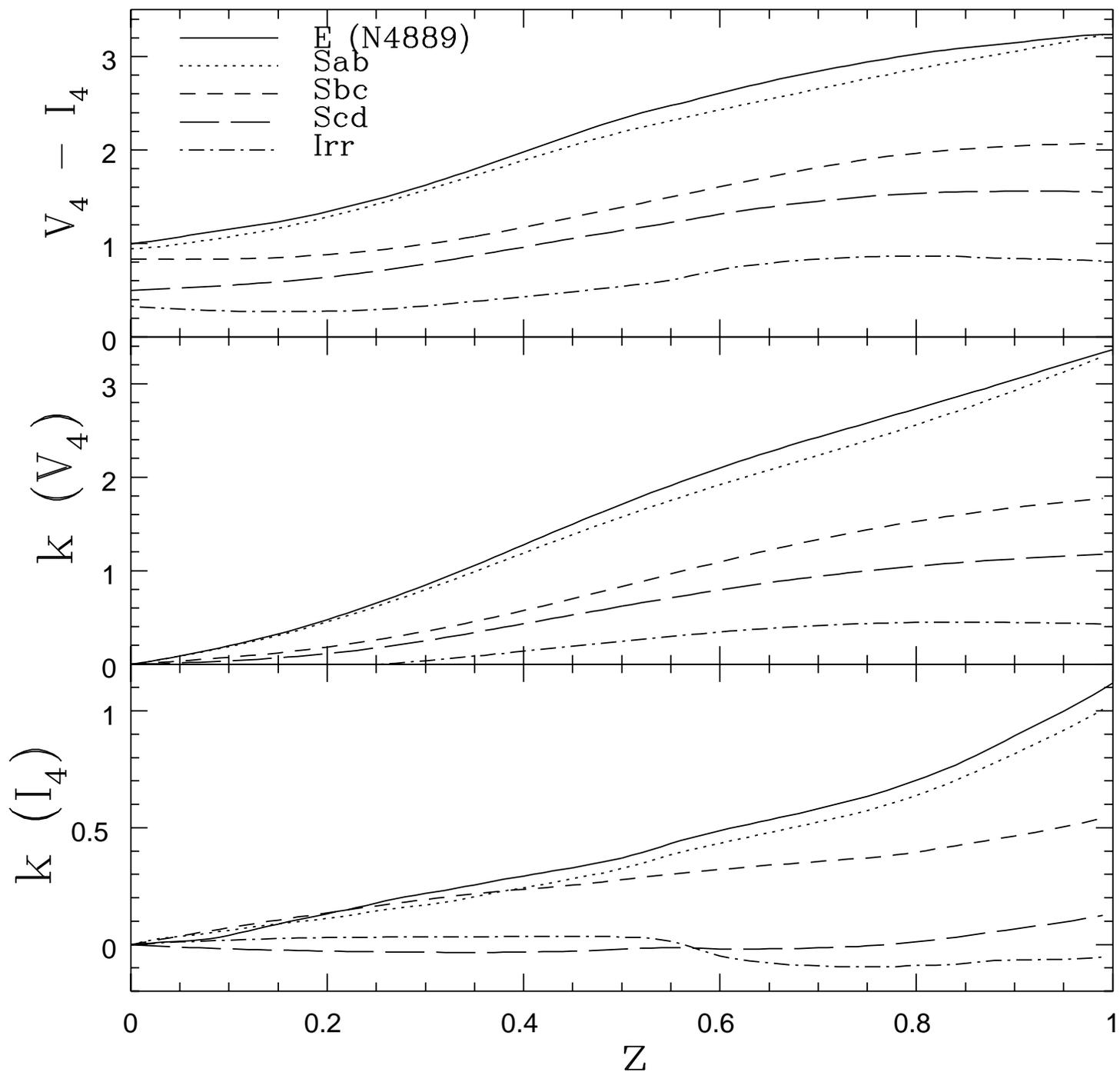}}
\caption{The no-evolution $V_{4} - I_{4}$ color and the \v4 and \i4 band
k-corrections as a function of redshift and morphological type as
calculated using the spectral energy distributions from Coleman, Wu \&
Weedman (1980).}
\end{figure}

\begin{figure}
\centerline{
\epsfysize=7.5in
\epsfbox{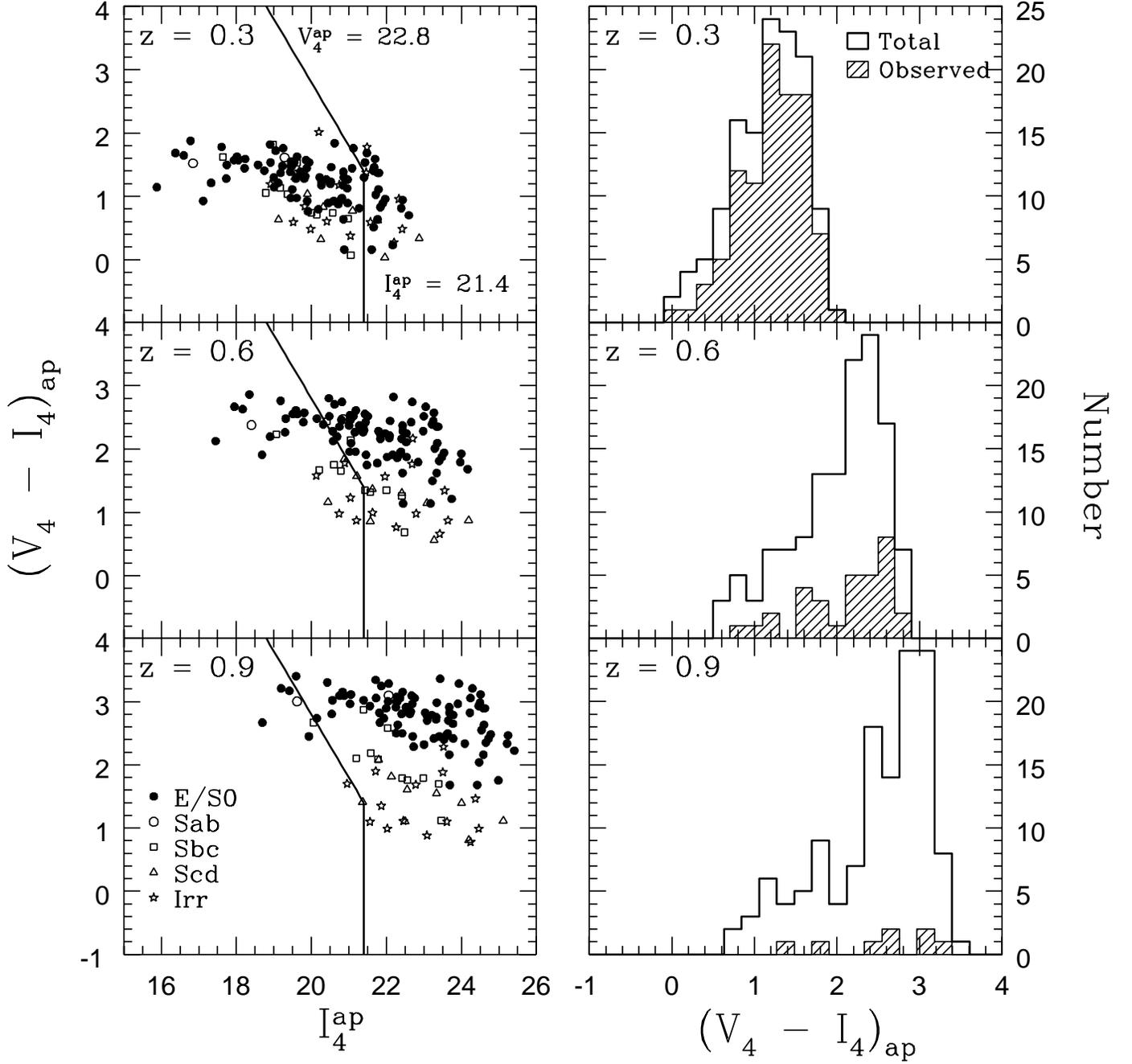}}
\caption{Left : the expected color--magnitude relation for a simulated
no--evolution cluster population as a function of redshift (see \S
3.1). The cluster redshift is indicated in the upper left corner of
each panel. E/S0s are indicated by filled circles. Sab, Sbc, Scd, and
Irr galaxies are indicated by circles, squares, triangles, and stars,
respectively. The magnitude limits of the PDCS ($V_{4}^{ap} \approx
22.8$ and $I_{4}^{ap} \approx 21.4$) are indicated by solid
lines. Right : the resulting color distributions. Solid line
histograms represent the total color distribution, while shaded
histograms represent the color distributions that would be observed in
our survey.}
\end{figure}

\begin{figure}
\centerline{\epsfbox{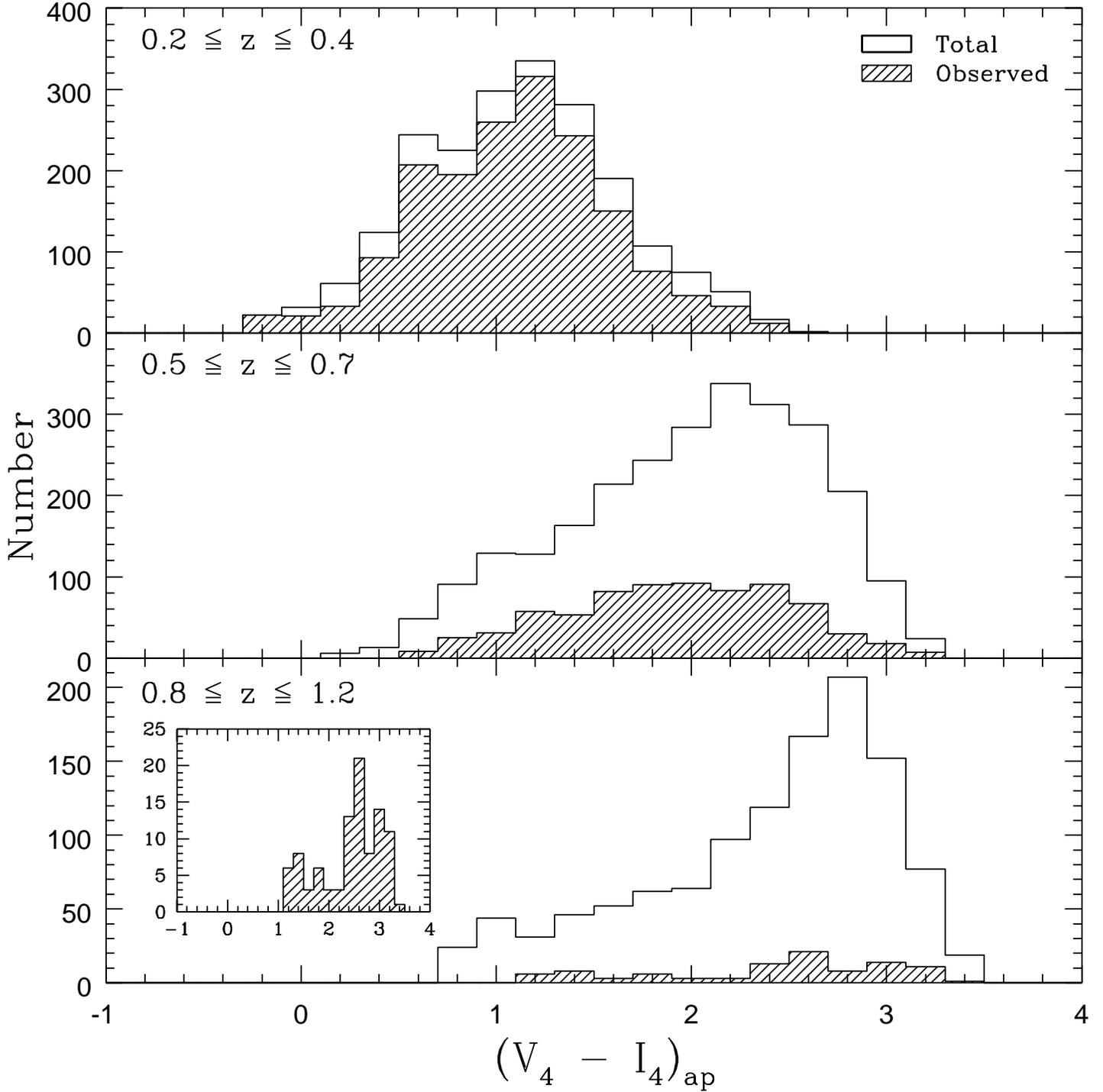}}
\caption{The expected composite color distributions in each of the
three redshift intervals for the simulated no--evolution cluster
populations (see \S 3.1).  Solid line histograms represent the total
color distribution, while shaded histograms represent the color
distribution that would be observed in our survey. For clarity, the
{\it observed} color distribution in the highest redshift interval is
expanded in the small window in the bottom panel.}
\end{figure}

\begin{figure}
\centerline{
\epsfysize=7.5in
\epsfbox{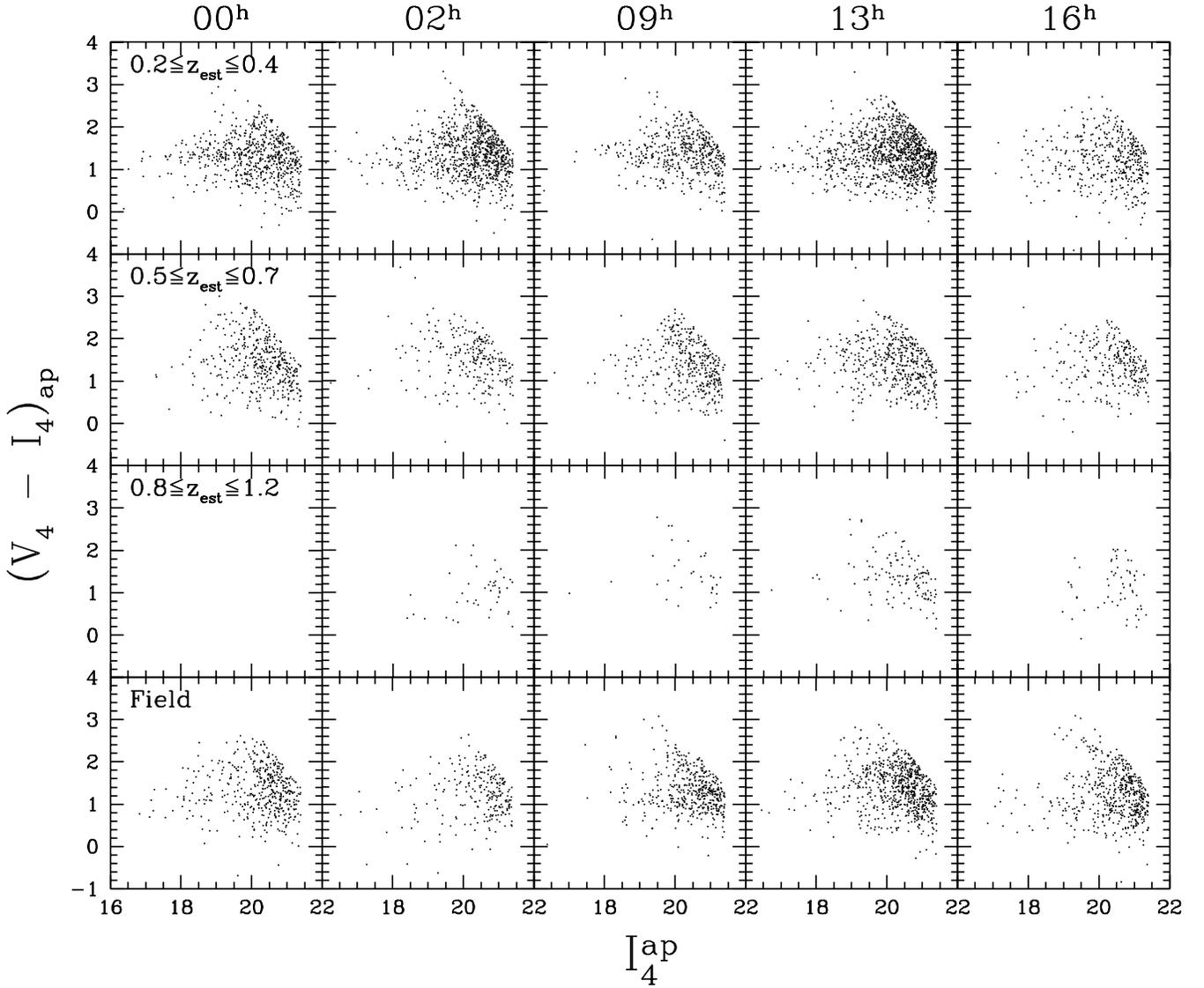}}
\caption{Composite color-magnitude diagrams for each of the five PDCS
fields. The aperture color, $(V_{4} - I_{4})_{ap}$, is plotted against
the aperture \i4 magnitude, $I_{4}^{ap}$. The aperture magnitude
limits have been applied. The top three panels show the composite CM
diagrams for each of the three redshift intervals (see \S3.2).  Table
2 lists the number of cluster candidates in each panel. The bottom
panels show the CM diagrams of regions of the five fields which
contain no cluster galaxies (indicated as ``field'').}
\end{figure}

\begin{figure}
\centerline{
\epsfysize=7.5in
\epsfbox{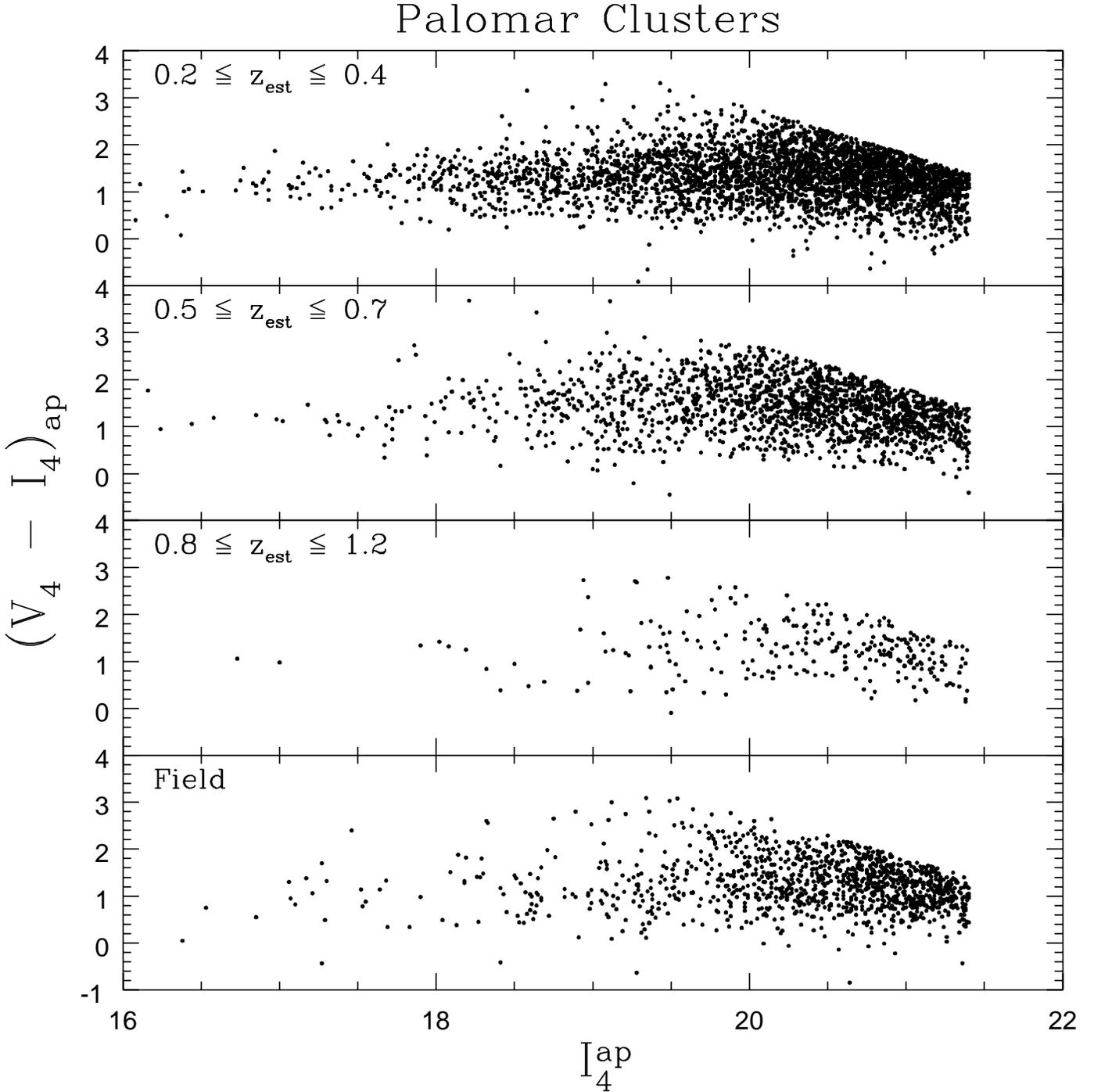}}
\caption{Composite color--magnitude diagrams of the Palomar clusters
in each of the three redshift intervals. The aperture color, $(V_{4} -
I_{4})_{ap}$, is plotted against the aperture \i4 magnitude,
$I_{4}^{ap}$. The aperture magnitude limits have been applied.  The
bottom panel shows the CM diagram of sample regions of the five
fields which contain no cluster galaxies (indicated as ``field'').}
\end{figure}

\begin{figure}
\centerline{
\epsfysize=7.5in
\epsfbox{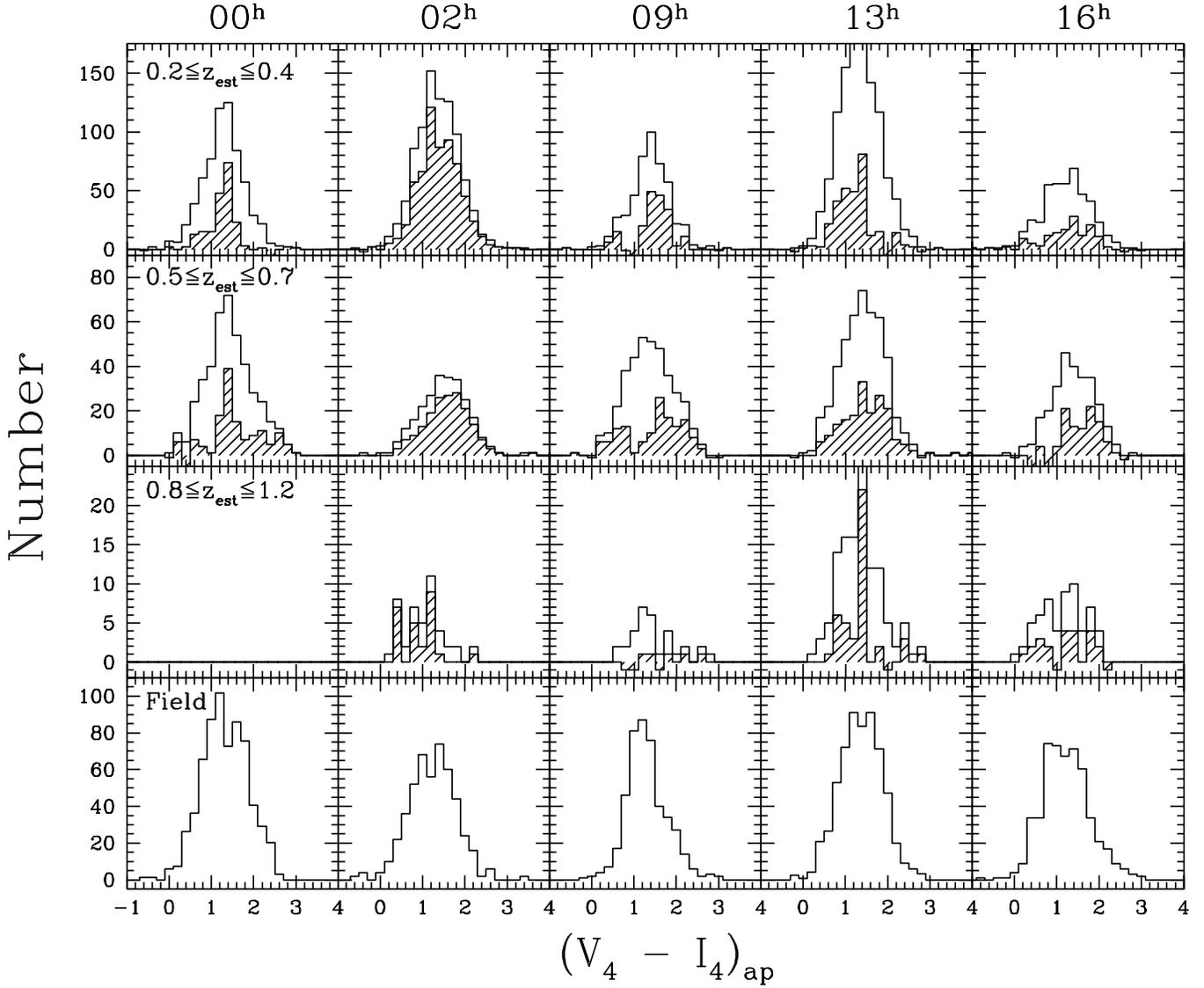}}
\caption{Composite aperture color distributions for each of the five
PDCS fields. The cluster galaxy color distributions are shown before
(solid line histograms) and after (shaded histograms) a statistical
correction for the background. The aperture magnitude limits have been
applied. The top three panels show the composite CM diagrams for each
of the three redshift intervals (see \S3.2).  The bottom panels show
the CM diagrams of regions of the five fields which contain no cluster
galaxies (indicated as ``field'').}
\end{figure}

\begin{figure}
\centerline{
\epsfysize=7.5in
\epsfbox{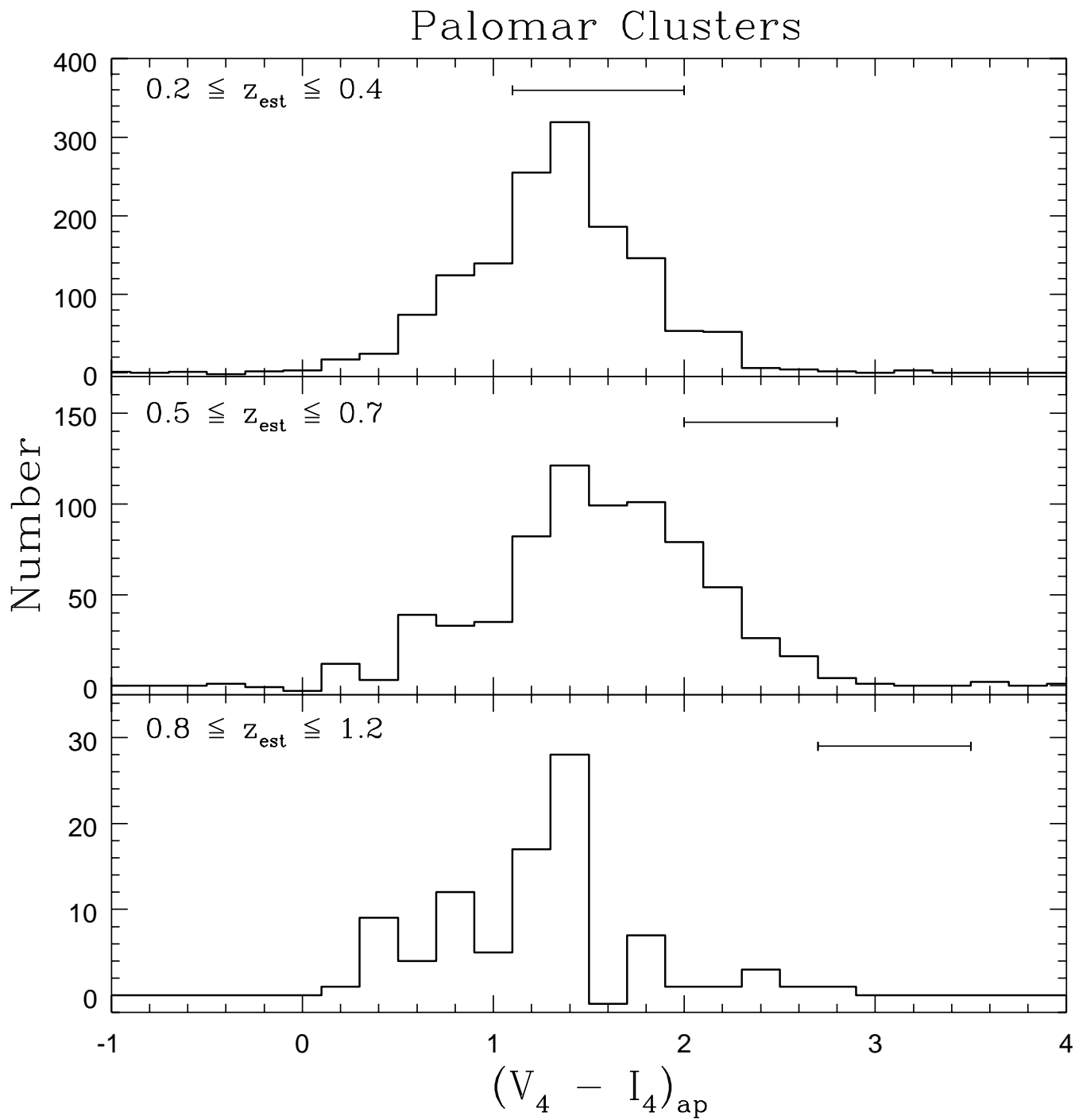}}
\caption{Composite aperture color distributions of the Palomar
clusters in each of the three redshift intervals.  Background has been
subtracted. The redshift interval is listed in the upper left of each
panel. The expected range of {\it no--evolution} E/S0s colors (Fig.\
2) for each redshift interval is indicated by a bar (see \S 3.2).}
\end{figure}

\begin{figure}
\centerline{
\epsfysize=6.8in
\epsfbox{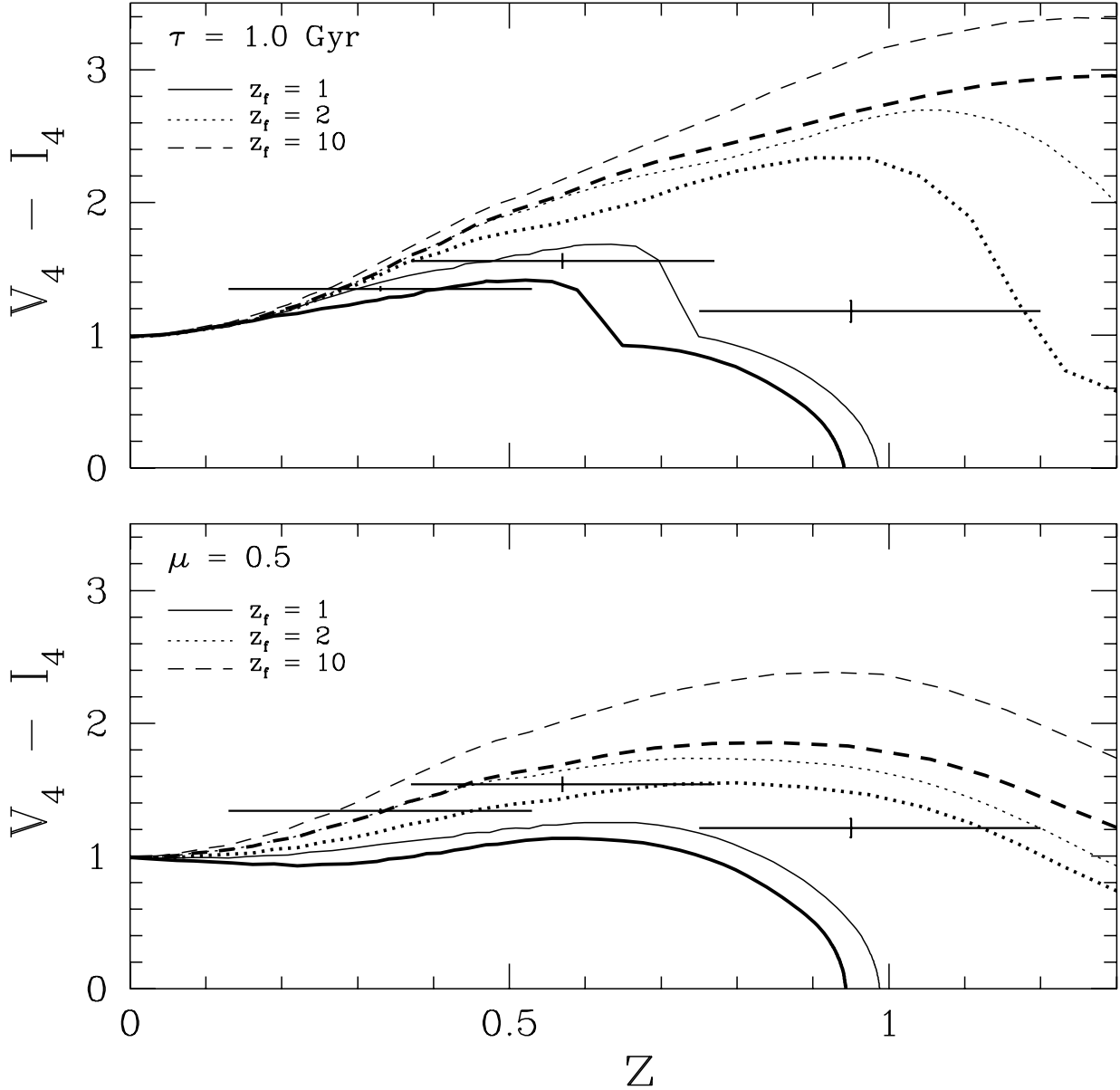}}
\caption{Comparison of the characteristic colors of the Palomar
clusters for each of the three redshift bins (see \S 3) with the
results of the Bruzual \& Charlot models (see \S 3.3) : a $\tau =
1~{\rm Gyr}$ burst of star formation (upper panel) and an
exponentially decaying star formation rate with $\mu = 0.5$ (lower
panel). The lines represent different epochs of the initial star
formation : $z_{f} = 1$ (solid lines), $z_{f} = 2$ (dotted lines), and
$z_{f} = 10$ (dashed lines) with $q_{o} = 0$ (thin lines) and $q_{o} =
0.5$ (thick lines). The vertical errors indicate the $1 \sigma$
confidence limits on the characteristic color of the
background-subtracted galaxy distribution (see \S 3.1). The horizontal
errors indicate the approximate range of estimated redshifts in each
bin.}
\end{figure}

\begin{figure}
\centerline{
\epsfysize=7.5in
\epsfbox{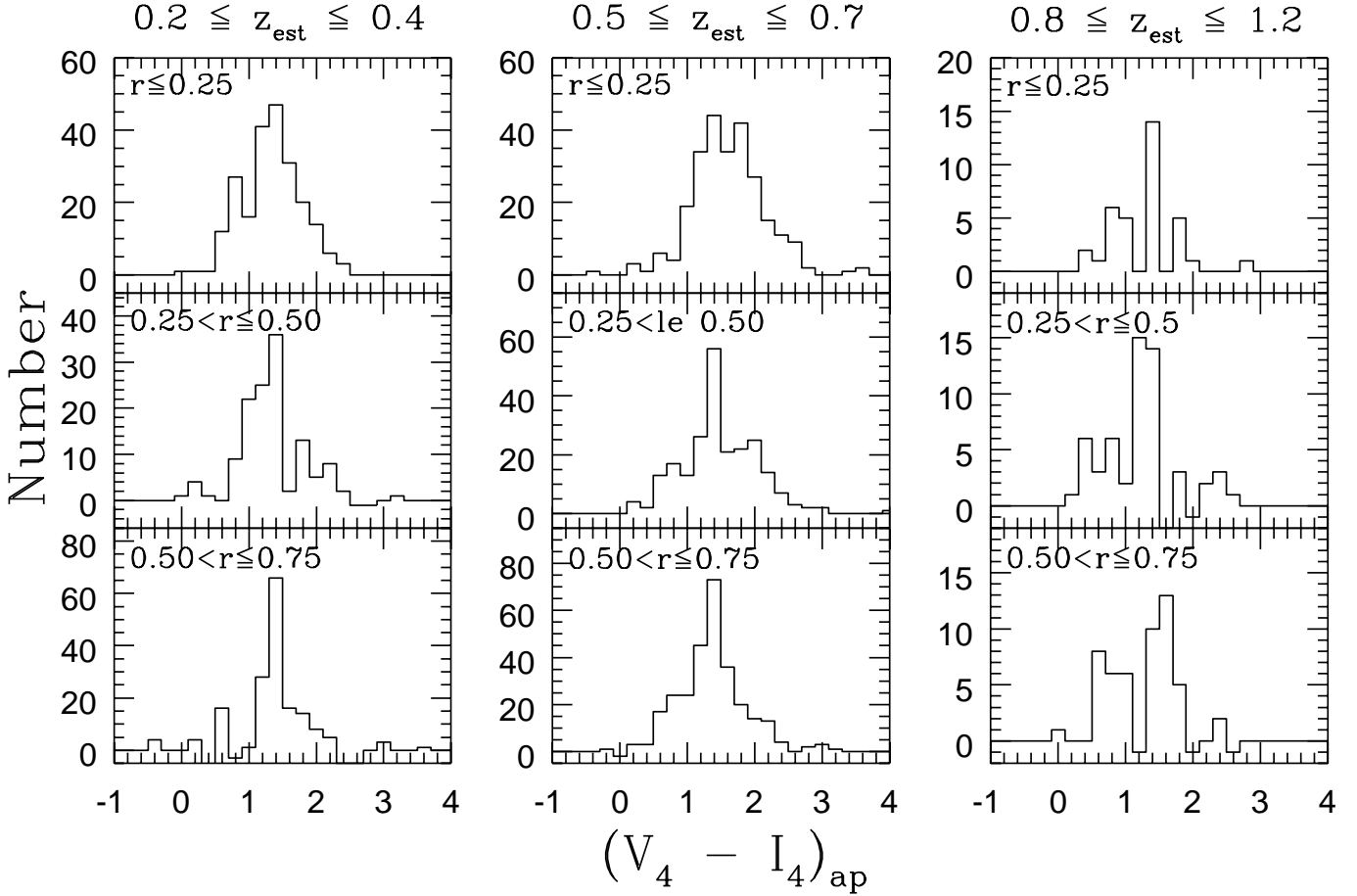}}
\caption{Composite aperture color distributions versus radial
distance for a richer ($\Lambda_{cl} \ge 70$) sample of the PDCS
clusters (see \S 3.4).  Columns indicate each of the three redshift
intervals.  The radial range (in $h^{-1}~{\rm Mpc}$) around the
cluster center is indicated in the upper left-hand corner of each
panel.}
\end{figure}

\begin{figure} 
\centerline{ 
\epsfysize=7.5in
\epsfbox{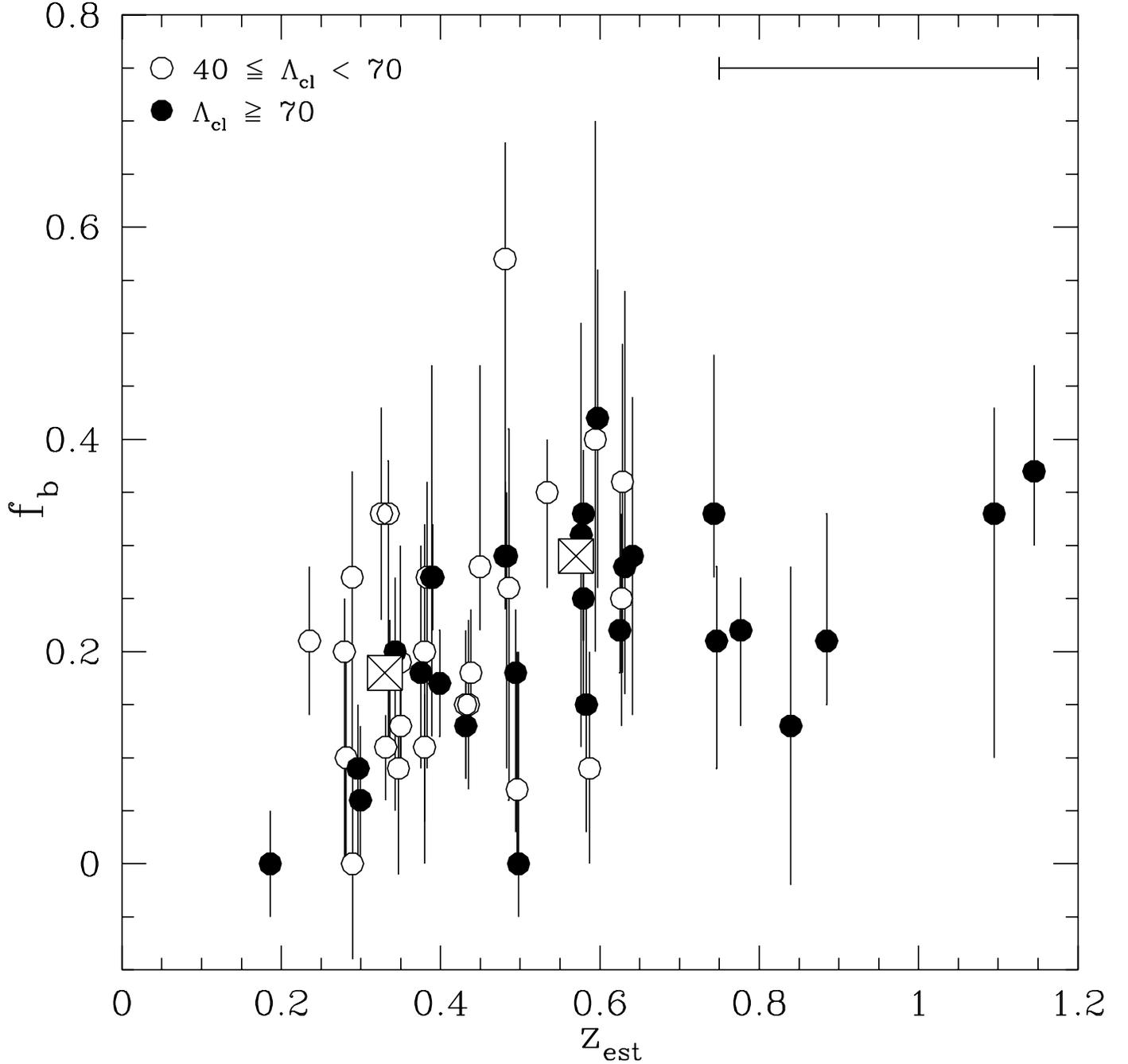}} 
\caption{The fraction of blue galaxies ($f_{b}$) versus estimated
redshift ($z_{est}$) for the 57 cluster candidates. In order to show
all points, we have randomly offset the cluster $z_{est}$ by less than
$\pm 0.05$.  A richer subsample of clusters ($\Lambda_{cl} \ge 70$) is
indicated by the filled circles. $f_{b}$ is only complete to $z_{est}
\approx 0.6$. The errors in $f_{b}$ represent the range of possible
values. The estimated redshift uncertainty is indicated in the upper
right hand corner. The median values in the first two redshift bins
are designated by the large boxed crosses. $f_{b}$ and $z_{est}$ are
correlated at a 96.2\% ($\sim 2 \sigma$) confidence level (see \S
4). }
\end{figure}

\begin{figure}
\centerline{
\epsfysize=7.5in
\epsfbox{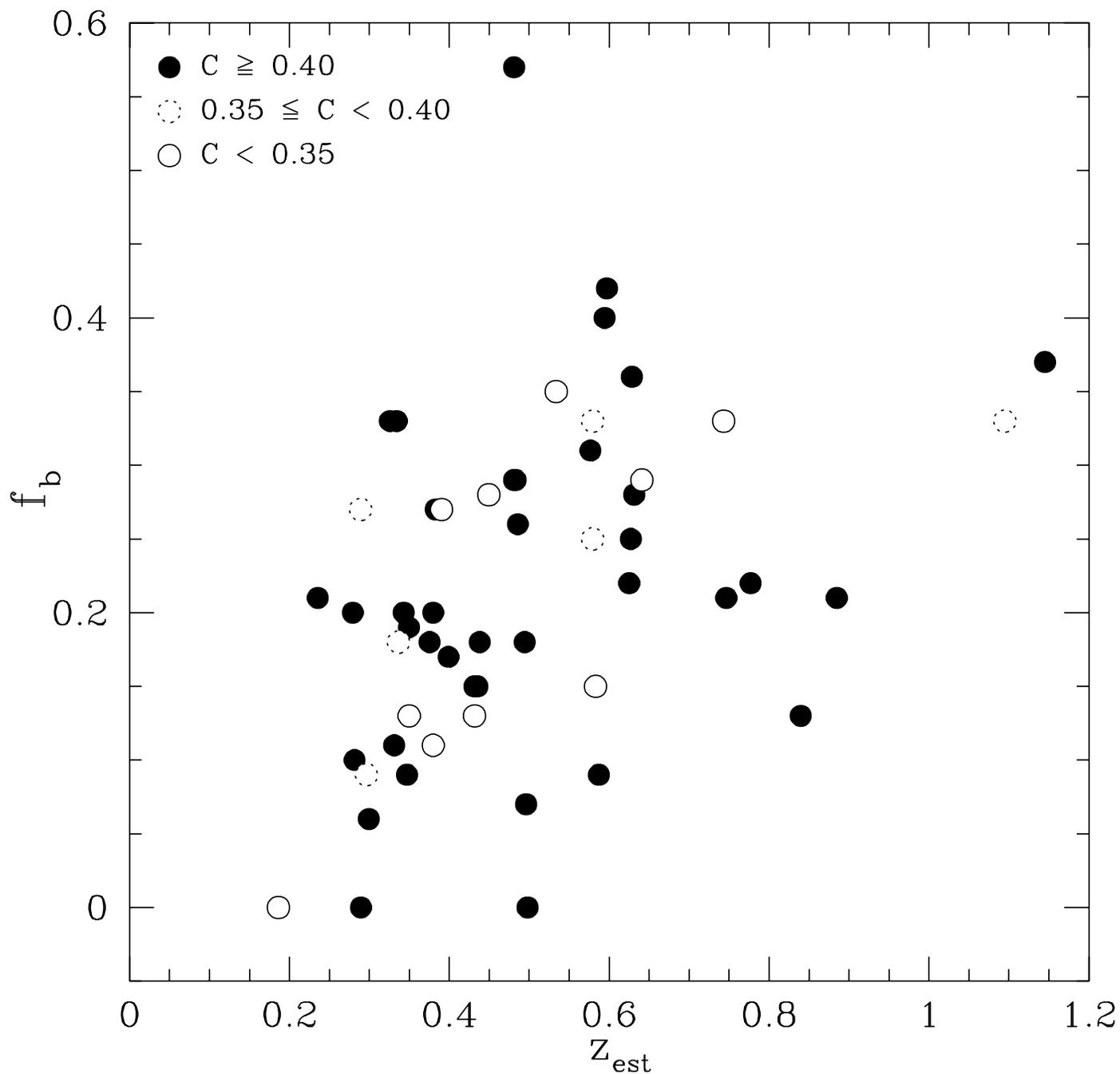}}
\caption{The fraction of blue galaxies ($f_{b}$) versus estimated
redshift for the 57 cluster candidates. The symbols indicate three
compactness ($C$) ranges of the cluster candidates (see \S 4). Filled
circles, compact clusters ($C \ge 0.4$); open circles, open clusters
($C < 0.35$); dotted circles, intermediate clusters ($0.35 \le C <
0.4$).}
\end{figure}

%
%

\end{document}